\documentclass[%
 reprint,
superscriptaddress,
frontmatterverbose, 
reprint,
 amsmath,amssymb,
 aps,
prd,
floatfix,
]{revtex4-1}

\usepackage{graphicx}
\usepackage{dcolumn}
\usepackage{bm}
\usepackage[mathlines]{lineno}

\usepackage{color}
\usepackage{appendix}

\usepackage{romannum}

\begin{document}


\title{Nature of the ultrarelativistic prompt emission phase of GRB 190114C}

\author{R.~Moradi}
\affiliation{ICRA and Dipartimento di Fisica, Universit\`a  di Roma ``La Sapienza'', Piazzale Aldo Moro 5, I-00185 Roma, Italy}
\affiliation{International Center for Relativistic Astrophysics Network, Piazza della Repubblica 10, I-65122 Pescara, Italy}
\affiliation{INAF -- Osservatorio Astronomico d'Abruzzo,Via M. Maggini snc, I-64100, Teramo, Italy.}

\email{ rahim.moradi@inaf.it}

\author{J.~A.~Rueda}
\affiliation{ICRA and Dipartimento di Fisica, Universit\`a  di Roma ``La Sapienza'', Piazzale Aldo Moro 5, I-00185 Roma, Italy}
\affiliation{International Center for Relativistic Astrophysics Network, Piazza della Repubblica 10, I-65122 Pescara, Italy}
\affiliation{ICRANet-Ferrara, Dipartimento di Fisica e Scienze della Terra, Universit\`a degli Studi di Ferrara, Via Saragat 1, I--44122 Ferrara, Italy}
\affiliation{Dipartimento di Fisica e Scienze della Terra, Universit\`a degli Studi di Ferrara, Via Saragat 1, I--44122 Ferrara, Italy}
\affiliation{INAF, Istituto di Astrofisica e Planetologia Spaziali, Via Fosso del Cavaliere 100, 00133 Rome, Italy}
\email{jorge.rueda@icra.it}

\author{R.~Ruffini}
\affiliation{ICRA and Dipartimento di Fisica, Universit\`a  di Roma ``La Sapienza'', Piazzale Aldo Moro 5, I-00185 Roma, Italy}
\affiliation{International Center for Relativistic Astrophysics Network, Piazza della Repubblica 10, I-65122 Pescara, Italy}
\affiliation{INAF, Viale del Parco Mellini 84, 00136 Rome, Italy}
\email{ ruffini@icra.it}
\author{Liang~Li}
\affiliation{ICRA and Dipartimento di Fisica, Universit\`a  di Roma ``La Sapienza'', Piazzale Aldo Moro 5, I-00185 Roma, Italy}
\affiliation{International Center for Relativistic Astrophysics Network, Piazza della Repubblica 10, I-65122 Pescara, Italy}
\affiliation{INAF, Viale del Parco Mellini 84, 00136 Rome, Italy}
\email{ liang.li@icranet.org}

\author{C.~L.~Bianco}
\affiliation{ICRA and Dipartimento di Fisica, Universit\`a  di Roma ``La Sapienza'', Piazzale Aldo Moro 5, I-00185 Roma, Italy}
\affiliation{International Center for Relativistic Astrophysics Network, Piazza della Repubblica 10, I-65122 Pescara, Italy}
\affiliation{INAF, Istituto di Astrofisica e Planetologia Spaziali, Via Fosso del Cavaliere 100, 00133 Rome, Italy}

\author{S.~Campion}
\affiliation{ICRA and Dipartimento di Fisica, Universit\`a  di Roma ``La Sapienza'', Piazzale Aldo Moro 5, I-00185 Roma, Italy}
\affiliation{International Center for Relativistic Astrophysics Network, Piazza della Repubblica 10, I-65122 Pescara, Italy}

\author{C.~Cherubini }
\affiliation{International Center for Relativistic Astrophysics Network, Piazza della Repubblica 10, I-65122 Pescara, Italy}
\affiliation{ICRA, University Campus Bio-Medico of Rome, Via Alvaro del Portillo 21, I-00128 Rome, Italy}
\affiliation{Department of Science and Technology for Humans and the Environment and  Nonlinear Physics and Mathematical Modeling Lab, \\University Campus Bio-Medico of Rome, Via Alvaro del Portillo 21, 00128 Rome, Italy}

\author{S.~Filippi}
\affiliation{International Center for Relativistic Astrophysics Network, Piazza della Repubblica 10, I-65122 Pescara, Italy}
\affiliation{ICRA, University Campus Bio-Medico of Rome, Via Alvaro del Portillo 21, I-00128 Rome, Italy}
\affiliation{Department of Engineering, University Campus Bio-Medico of Rome, Nonlinear Physics and Mathematical Modeling Lab, \\Via Alvaro del Portillo 21, 00128 Rome, Italy}
\author{Y.~Wang}
\affiliation{ICRA and Dipartimento di Fisica, Universit\`a  di Roma ``La Sapienza'', Piazzale Aldo Moro 5, I-00185 Roma, Italy}
\affiliation{International Center for Relativistic Astrophysics Network, Piazza della Repubblica 10, I-65122 Pescara, Italy}
\affiliation{INAF -- Osservatorio Astronomico d'Abruzzo,Via M. Maggini snc, I-64100, Teramo, Italy}
\email{ yu.wang@uniroma1.it}
\email{ liang.li@icranet.org}
\author{S.~S.~Xue}
\affiliation{ICRA and Dipartimento di Fisica, Universit\`a  di Roma ``La Sapienza'', Piazzale Aldo Moro 5, I-00185 Roma, Italy}
\affiliation{International Center for Relativistic Astrophysics Network, Piazza della Repubblica 10, I-65122 Pescara, Italy}
\affiliation{}

\date{\today / Received date /
Accepted date }

\begin{abstract}
We address the physical origin of the ultrarelativistic prompt emission (UPE) phase of GRB 190114C observed in the interval $t_{\rm rf}=1.9$--$3.99$~s, by the \textit{Fermi}-GBM in 10 keV--10 MeV energy band. Thanks to the high signal–to–noise ratio of \textit{Fermi}-GBM data, \textcolor{black}{a time-resolved spectral analysis has evidenced a sequence of similar blackbody plus cutoff power-law spectra (BB+CPL), on ever decreasing time intervals during the entire UPE phase.} We assume that  during  the UPE phase, the ``\emph{inner engine}'' of the GRB, \textcolor{black}{composed of a Kerr black hole (BH) and a uniform test magnetic field $B_0$, aligned with the BH rotation axis,} operates in an overcritical field  $|{\bf E}|\geq E_c$, where $E_c=m_e^2 c^3/(e\hbar)$, being $m_e$ and $-e$ the mass and charge of the electron. \textcolor{black}{We infer an $e^+~e^-$ pair electromagnetic plasma in presence of a baryon load, \emph{a PEMB pulse},  originating from a vacuum polarization quantum process in the \textit{inner engine}.} This initially optically thick plasma self-accelerates, giving rise at the transparency radius to the MeV radiation observed by \textit{Fermi}-GBM. At times $t_{\rm rf}>3.99$~s, the electric field becomes undercritical, $|{\bf E}|<E_c$, and the \textit{inner engine}, \textcolor{black}{as previously demonstrated}, operates in the  classical electrodynamics regime and generate the GeV emission. \textcolor{black}{ During both the ``quantum'' and the ``classical'' electrodynamics processes, we determine the time varying mass and spin of the Kerr BH in the \textit{inner engine},} fulfilling the Christodoulou-Hawking-Ruffini mass-energy formula of a Kerr BH.  \textcolor{black}{For the first time, we quantitatively show how the \textit{inner engine}, by extracting the rotational energy of the Kerr BH, produces a series of PEMB pulses. We follow the quantum vacuum polarization process in sequences with decreasing time bins. We compute the Lorentz factors, the baryon loads and the radii at transparency, as well as the value  of the magnetic field, $B_0$, assumed to be constant in each sequence. The fundamental hierarchical structure, linking the quantum electrodynamics regime to the classical electrodynamics regime, is characterized by the emission of ``\textit{blackholic quanta}'' with a} timescale $\tau \sim 10^{-9}$~s, and energy $\mathcal{E} \sim 10^{45}$~erg. 

\end{abstract}
\maketitle

\section{Introduction}\label{sec:1}

\textcolor{black}{It is by now clear that gamma-ray bursts (GRBs), far from being a short single elementary process lasting less than 10$^2$~s, are possibly the most complex astrophysical systems in the Universe, an authentic fundamental physics laboratory. A series of Episodes, corresponding to new different physical laws, take place on vastly different characteristic time scales ranging from quantum electrodynamics (QED) time scales of $10^{-21}$~s, to classical electrodynamics processes of $10^{-14}$~s, as well as to gravitational processes of $10^{-6}$ s, and to hydrodynamics timescales of $1$ s and of $10^7$~s, and the GRB source lifetime can indeed be as long as $10^{17}$~s, i.e. the Universe lifetime (see \cite{2021MNRAS.tmp..868R}, and references therein).}

\textcolor{black}{One of the most intriguing phenomena occurring in the most energetic long GRBs is the ultrarelativistic prompt emission (UPE) phase. In the case of GRB 190114C: 1) it takes place on a $2$~s rest-frame time ($t_{\rm rf}$) interval, 2) it encompasses $40\%$ of the GRB isotropic energy, and 3) it occurs in an originally optically thick domain reaching transparency in the keV-MeV energy range.}

\textcolor{black}{We address in this article the challenge of inferring, from spectral properties, on an ever-decreasing time scales, the nature of this new and yet unexplained process. We use:
A) The concepts previously developed for a self-accelerating optically thick $e^+~e^-$ pair-electromagnetic-baryon plasma (PEMB pulse) originated from vacuum polarization produced by a overcritically charged Kerr-Newman black hole (BH) \cite{1975PhRvL..35..463D, 1998astro.ph.11232R,PhysRevD.17.1518,1999A&A...350..334R,2000A&A...359..855R,2010PhR...487....1R}.
%
B) A specific property of the PEMB pulse:  the reaching of the transparency radius with $\Gamma \sim $ 100 \citep{1998astro.ph.11232R}, which is essential to overcome the compactness problem
of the UPE phase; see e.g. \citep[][]{1975NYASA.262..164R,2004RvMP...76.1143P}
C) The Papapetrou-Wald solution \citep{1966AIHPA...4...83P,1974PhRvD..10.1680W} as an alternative to the charged Kerr-Newman BH as the source of vacuum polarization process. 
D) The concept of an ``effective charge'', $Q_{\rm eff}$, which overcomes the difficulty of adopting the unexplained origin of a charged BH. This concept allows to explain in an ``effective way'' the electric field which arises from the gravitomagnetic interaction of a Kerr BH with a surrounding magnetic field, $B_0$. The effective charge has been used in the study of the ``\emph{inner engine}'' of GRB 130427A \cite{2019ApJ...886...82R} and GRB 190114C \cite{2021A&A...649A..75M}.}

\textcolor{black}{We address the UPE phase of GRB 190114C that, owing to its morphology, can be identified as a canonical binary driven hypernova (BdHN) of type I (see details below), observed with a viewing angle orthogonal to the orbital plane of the GRB binary progenitor. Indeed, a variety of Episodes of GRB 190114C have been already identified and duly explained \citep{2021MNRAS.tmp..868R}, including the X-ray afterglow \citep{2020ApJ...893..148R} and the GeV emission \citep{2021A&A...649A..75M}. }

It has been possible since the beginning of 2018 \citep{2018ApJ...869..101R,2019ApJ...874...39W,2020ApJ...893..148R,2020EPJC...80..300R,2021MNRAS.tmp..868R} to obtain specific new results thanks to a variety of factors, including the identification of new \textcolor{black}{GRB} paradigms, a novel time-resolved spectral analysis fulfilling stringent criteria of statistical significance \citep{Li2019,Li2019a,Li2019b,Li2020,Li2020b,Li2021}, and three-dimensional, numerical smoothed-particle-hydrodynamic (SPH) simulations of BdHNe presented in \citet{2019ApJ...871...14B}. From these results, it has been concluded:
 
\textcolor{black}{A) There is clear} evidence that the progenitors of long GRBs are binary systems composed of a \textcolor{black}{carbon-oxygen (CO)} star and a neutron star (NS) \textcolor{black}{companion: the BdHN}. \textcolor{black}{The gravitational collapse of the iron core of the CO star leads to the SN and forms the newborn NS  ($\nu$NS).} \textcolor{black}{When the binary period is short i.e. $\sim$ 5 min,} the SN ejecta hypercritically accretes onto the companion NS, leading to the formation of a BH \citep{,2021MNRAS.tmp..868R}. \textcolor{black}{These systems are known as  BdHN of type I (BdHN I).} \textcolor{black}{This approach was successfully adopted in explaining} the physical origin of the X-ray flares \citep{2018ApJ...852...53R}, further confirmed in \citet{2018ApJ...869..151R}. 
 
\textcolor{black}{B) The accretion of the SN ejecta onto the $\nu$NS} in BdHNe has given the opportunity to explain the underlying physical nature of the \textcolor{black}{X-ray} afterglow in GRB 130427A, GRB 160509A, GRB 160625B, GRB 180720A and GRB 190114C; see \cite{2018ApJ...869..101R} and \cite{2020ApJ...893..148R}.

\textcolor{black}{C) T}he observations of a mildly relativistic phase in the GRB plateau and in the afterglow \citep{2018ApJ...852...53R, 2018ApJ...869..151R}, have motivated the use of BdHNe model in order to explain the energetic of the GeV emission as originating from the extraction of rotational energy of a Kerr BH very close to the BH horizon, \textcolor{black}{described by the \textit{inner engine}}, addressed in \citet{2019ApJ...886...82R,2020EPJC...80..300R}.

\textcolor{black}{The \textit{inner engine} is composed of: 1) a Kerr BH with mass of $M$ and angular momentum of $J$, 2) an asymptotically uniform magnetic field, $B_0$ aligned with the BH rotation axis, the Papapetrou-Wald solution \citep{1974PhRvD..10.1680W}, and 3) a very low density plasma composed of ions and electrons with density of $10^{-14}$ g cm$^{-3}$; see \citet{2019ApJ...886...82R}. 
The effective charge, $Q_{\rm eff}$, of this system: }
\begin{eqnarray}\label{eq:EFCH}
  \textcolor{black}{Q_{\rm eff}=2 B_0 J G/c^3,}
\end{eqnarray} 
\textcolor{black}{originates from the gravitomagnetic interaction of the Kerr BH with the surrounding magnetic field, $B_0$, being $c$ and $G$ the speed of light in vacuum and the gravitational constant, respectively; see \citet{2019ApJ...886...82R}, \citet{2020EPJC...80..300R} and \citet{2021A&A...649A..75M}.}

\textcolor{black}{In order to explain the GeV emission the \emph{inner engine} operates in an undercritical electric field regime, i.e. $|{\bf E}|< E_c$, where $E_c=m_e^2 c^3/(e\hbar)$, being $m_e$ and $-e$ the electron mass and charge, respectively, in presence of a magnetic field of $B_0\sim 10^{10}$~G, assumed to be constant in the entire process of emission. During this process:}
\begin{enumerate}
    \item
    \textcolor{black}{electrons are injected close to the horizon with an initial Lorentz factor of $\gamma=1$;}
    \item 
    \textcolor{black}{electrons are accelerated by the electromagnetic fields of the \emph{inner engine} and radiate synchrotron photons of GeV energies;}
    \item 
    \textcolor{black}{the radiation does not occur continuously but is emitted in elementary events (quanta) of $\sim 10^{37}$~erg on a time scale of $\sim 10^{-14}$~s. This energy is paid by the rotational energy of the Kerr BH implying a corresponding decrease of the angular momentum $J$ of the Kerr BH \citep{2019ApJ...886...82R, 2020EPJC...80..300R,2021A&A...649A..75M}. }
\end{enumerate}

 \textcolor{black}{The emission of the quanta is repetitive. After the emission of each quanta, a new process occurs starting from a new value $J^{*}=J-\Delta J$ of the angular momentum, with $\Delta J/J\sim 10^{-16}$, being $\Delta J$ the angular momentum extracted to the Kerr BH by the event in each repetitive step \citep{2019ApJ...886...82R, 2021A&A...649A..75M}.}

In this article, we \textcolor{black}{address the study of the UPE phase utilizing our previous background and being guided by: 
1) the hierarchical structure of the UPE in GRB 190114C with characteristic spectral signature of a cutoff power-law and a blackbody component (CPL$+$BB); see \citet{2019arXiv190404162R};}
\textcolor{black}{2) the \emph{inner engine} model which has been already well tested for the GeV radiation in GRB 130427A \citep{2019ApJ...886...82R} and GRB 190114C \citep{2021A&A...649A..75M}.} 

\textcolor{black}{We recall that the electric field in the Papapetrou-Wald solution in the slow-rotation approximation is given by \citep{2019ApJ...886...82R}: }
\begin{equation}\label{eq:ER22}
  \textcolor{black}{E_r \approx -\frac{1}{2}\alpha B_0\,\frac{r_+^2}{r^2},}
\end{equation}
\textcolor{black}{where $r_+$ is the outer event horizon and $\alpha \equiv cJ/(GM^2)$ is the dimensionless BH spin parameter.}

\textcolor{black}{The profound novelty characterizing the UPE phase is the assumption of an overcritical field, i.e., $|{\bf E}|\geq E_c$ around the Kerr BH in the \emph{inner engine}. This overcritical field generates, via vacuum polarization, an optically thick PEMB pulse which owing to its high density (here $\sim 10^{8}$ g cm$^{-3}$) and high interior pressure, self-accelerates to an ultrarelativistic regime and finally reaching the transparency point \citep{2010PhR...487....1R}.} 

\textcolor{black}{The hydrodynamic equations of the relativistic expanding PEMB pulses are integrated until the point of transparency when the MeV radiation becomes observable. The radius of transparency and Lorentz factor are explicitly evaluated. This solution was first addressed in \citet{1998astro.ph.11232R,1999A&A...350..334R,2000A&A...359..855R,2010PhR...487....1R}. This is the fundamental physical process which is assumed to be at the very ground of the description of the UPE phase and its spectral properties. Again, the energy in the overcritical field originates from the rotational energy of the Kerr BH in the Papapetrou-Wald solution.}

\textcolor{black}{An additional necessary step is how to carry out the matching of the overcritical regime, characterizing the UPE phase, its MeV radiation, and its intrinsic quantum nature, with the already analysed undercritical regime following the UPE phase.  This undercritical regime describes the GeV radiation and is dominated in the \emph{inner engine} by a classical electrodynamics nature with very low density surrounding plasma.}

\textcolor{black}{For the determination of the parameters of the \textit{inner engine}, we are guided by the time-resolved spectra analysis and existence of the hierarchical structure found in the UPE phase \citep{2019arXiv190404162R}. Each successive iteration (rebinning) fulfills the total energy requirement and spectral structures in different timescales. We select as the fundamental iterative process the ``only'' one which allows the electric field to fulfill at the end of the UPE phase the constraint $|{\bf E}|=E_c$. This boundary condition determines the value of $B_0$ and is necessary to join the UPE phase to the classical electrodynamics regime, originating the GeV radiation.}

\textcolor{black}{Similar to the case of the generation of GeV radiation from the \emph{inner engine}, also the emission of the MeV radiation during the UPE phase is not continuous:}
\begin{enumerate}
    \item \textcolor{black}{$e^+~e^-\gamma$ plasma, in presence of the baryon load, is generated by the vacuum polarization close to horizon with initial bulk Lorentz factor $\Gamma=1$ on a characteristic timescale of $\sim \hbar/(m_e c^2) \approx 10^{-21}~\rm s$,
    \item these PEMB pulses self-accelerate all the way to the point of transparency at which emit MeV radiation in an ultrarelativistic regime,
    \item the process is again repetitive; at the end of each step the process restarts with a value of electric field given by Eq.~(\ref{eq:ER22}), keeping the magnetic field constant, but with a new value of the BH dimensionless spin parameter $\alpha^{*}=\alpha-\Delta \alpha$, with $\Delta \alpha/\alpha\sim 10^{-9}$, being $\Delta \alpha$ the amount of dimensionless BH spin extracted to the Kerr BH by the event in each repetitive step.}
    
\end{enumerate}
\textcolor{black}{The UPE phase stops in the sequence which allows the condition  $|{\bf E}|= E_c$ to be reached at the right time.}

In Sec. \ref{sec:3}, we recall three different Episodes identified in time-resolved spectral analysis of GRB 190114C. We focus on the spectral analysis of the Fermi-GBM (keV-MeV) and the Fermi-LAT (GeV) data during and after the UPE phase. 

\textcolor{black}{In Sec. \ref{sec:heir}, we present the time-resolved analysis of the UPE phase as well as} the appearance of the hierarchical structure of its spectra. \textcolor{black}{These results were announced in \citet{2019arXiv190404162R} and here presented in an improved numerical analysis, with their theoretical modeling.}

In Sec. \ref{sec:5}, we outline \textcolor{black}{the properties of} the \emph{inner engine}. This is composed of a uniform magnetic field aligned with the rotation axis of a Kerr BH, following the exact, mathematical solution of the Einstein-Maxwell equations given by \citet{1974PhRvD..10.1680W}. We here apply this solution to the astrophysical conditions occurring in a BdHN I.

In Sec. \ref{sec:mass-spin}, for $t_{\rm rf} > 3.99$~s, namely after the UPE phase, following \citet{2019ApJ...886...82R,2021MNRAS.tmp..868R}, \citet{2020EPJC...80..300R} and \citet{2021A&A...649A..75M}, we proceed to the self-consistent determination of \textcolor{black}{a)} the mass and spin of the BH, \textcolor{black}{b)} the magnetic field $B_0$. These parameters are determined to fulfill the \textcolor{black}{energetics of GeV radiation and} its transparency \textcolor{black}{with respect to} the process of pair production by photon-magnetic field interaction. The mass and spin of BH at $t_{\rm rf}$=$3.99$~s are, respectively, $M=4.45 M_\odot$ and $\alpha = 0.41$, and magnetic field is $B_0\sim10^{10}$~G.

\textcolor{black}{I}n Sec. \ref{sec:massupe}, \textcolor{black}{we determine} the \textcolor{black}{mass and spin of BH at $t_{\rm rf}$=$1.9$~s, $M=4.53 M_{\odot}$ and $\alpha=0.54$}. \textcolor{black}{This result is consistent with the luminosity obtained from the time-resolved spectral analysis, during} the UPE phase, \textcolor{black}{and the above values of the mass and spin for $t_{\rm rf} >$ 3.99~s given in Sec. \ref{sec:mass-spin}; see Figs.~\ref{fig:lumupe} and \ref{massspinupe}}. 

In Sec. \ref{sec:vac}, we  address the  overcritical regime, $|{\bf E}|\geq E_c$, in order to have the vacuum polarization \textcolor{black}{via} Schwinger $e^+~e^-$  pair production \citep[see discussion in][]{2009PhRvD..79l4002C}, in the UPE phase\textcolor{black}{; see also \citet{1998astro.ph.11232R,1999A&A...350..334R,2000A&A...359..855R,2010PhR...487....1R}}.

In Sec.~\ref{sec:12}, we assess the general formulation of the transparency of the MeV photons during the UPE phase.  

In Sec.~\ref{sec:mag-fifth}, we determine the magnetic field, $B_0\sim 10^{17}~$G, inferred from the time-resolved spectral analysis \textcolor{black}{with $\Delta t=0.125$~s resolution, represented in Sec. \ref{sec:heir}}, \textcolor{black}{corresponding to the} emission at the transparency point of the $16$ \textcolor{black}{PEMB pulses with the repetition time} of $\tau= 0.125~$s. \textcolor{black}{This sequence does not fulfill the boundary condition of the UPE phase i.e., $|{\bf E}|= E_c$ at $t_{\rm rf}=3.99~$s; see Fig.~\ref{fig:Lrentz1051}}

In Sec.~\ref{sec:mag-low}, we obtain the lower limit of magnetic filed around the BH, $B_0= 2.3 \times 10^{14}~$G, during the UPE phase by imposing $|{\bf E}|= E_c$ at $t_{\rm rf}=3.99~$s, marking the end of UPE phase.  We infer that the UPE phase results from emission at the transparency point of the $\sim 10^9$ \textcolor{black}{PEMB pulses}, with radiation timescale of $\tau_q\sim 10^{-9}~$s\textcolor{black}{; see Fig.~\ref{fig:eduringupe}}.

\textcolor{black}{In Sec.~\ref{sec:comparison}, we make a comparison with other approaches}

In section \ref{sec:14}, we outline the conclusions of this article.
  
\section{\textit{Fermi} data of GRB 190114C}\label{sec:3}

%
\begin{figure*}
    \centering
\includegraphics[width=16 cm]{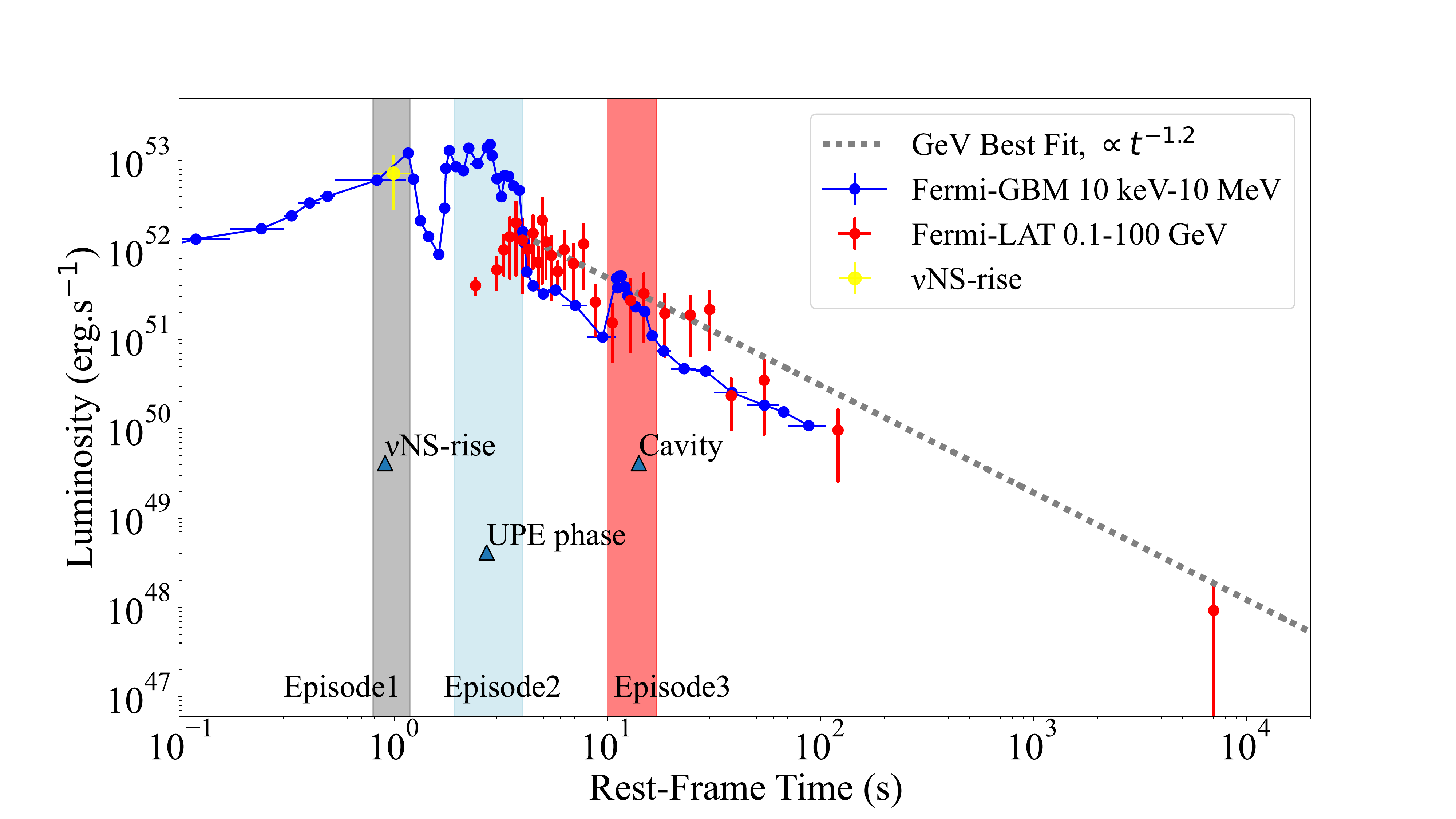}
     \caption{   Luminosity of the \textit{Fermi}-GBM in the $10$~keV--$10$~MeV energy band together with the luminosity of \textit{Fermi}-LAT during and after UPE phase expressed in the rest-frame of the source. The light grey part shows the \textcolor{black}{$\nu$NS}--rise from $t_{\rm rf}=$0.79~s to $t_{\rm rf}=$1.18~s. The light blue part shows the UPE phase which is in the time interval $t_{\rm rf}=1.9$--$3.99$~s, whose lower and upper edges correspond, respectively, to the moment of BH formation and to the moment which blackbody component disappears from the GBM data. The corresponding analysis for GRB  130427A, GRB 160509A and GRB 160625B is presented in \citep{2019arXiv191012615L}. The red part shows the Fermi-GBM the \textit{cavity} introduced in \citet{2019ApJ...883..191R}. The rest-frame $0.1$--$100$~GeV luminosity light-curve of GRB 190114C after UPE phase is best fitted by a power-law with slope of $1.2\pm 0.04$ and amplitude of $7.75 \times 10^{52}$~erg~s$^{-1}$}\label{fig:data}%
\end{figure*}
At 20:57:02.63 UT on 14 January 2019, \textit{Fermi}-GBM was triggered by GRB 190114C \citep{2019GCN.23707....1H}. The \textit{Fermi}-LAT had a boresight angle of $68$~degrees at the trigger time, the GRB remained in the field of view of \textit{Fermi}-LAT for $150$~s.   With the redshift of $z=0.424$ \citep{GCN23695} the isotropic energy of this burst is $E_{\rm iso}=(2.48 \pm 0.22) \times 10^{53}$erg. \textcolor{black}{Since BdHNe I are characterized by $E_{\rm iso}\gtrsim 10^{52}$~erg, we have identified GRB 190114C as a BdHN I and predict the occurrence of an associated SN \citep{2019GCN.23715....1R}}. \textcolor{black}{This prediction was followed by the successful observation}  of the SN associated with this burst \citep{2019GCN.23983....1G}. \textcolor{black}{The first GeV photon with probability more than 90$\%$ belonging to this GRB is a $\sim~0.9$~GeV photon observed at $t_{\rm rf}=1.9$~s after the GBM trigger.} The highest-energy photon is a $22.9$~GeV event which is observed $15$~s after the GBM trigger \citep{GCN23709}. GRB 190114C \textcolor{black}{has become since a prototype for identifying the BdHN I Episodes.}

\textcolor{black}{T}hree different Episodes \textcolor{black}{have been identified} in the \textit{Fermi}-GBM data; see Fig.~\ref{fig:data}:

Episode 1 with an isotropic energy of $E_{\rm iso}=(1.0 \pm 0.12) \times 10^{53}$~erg, occurs in the rest-frame time interval $t_{\rm rf}=[0, 1.9]$~s, being $t_{\rm rf}=0$~s the  \textit{Fermi}-GBM rest-frame trigger time. It reveals a thermal component from $t_{\rm rf}=0.79$~s to $t_{\rm rf}=1.18$~s, in its spectral analysis, marking the rise of \textcolor{black}{newly born NS} (\textit{\textcolor{black}{$\nu$NS}--rise}) with a corresponding isotropic energy of $E^{\rm\textcolor{black}{ iso}}_{\rm \textcolor{black}{\nu NS}}=(2.82 \pm 0.13) \times 10^{52}$~erg. 

Episode 2, with an isotropic energy of $E_{\rm iso}=(1.47 \pm 0.2) \times 10^{53}$~erg, \textcolor{black}{equivalent to 40\% of isotropic energy of the GRB}, lasts only $2$~s. It occurs in the rest-frame time interval $t_{\rm rf}=[1.9, 3.99]$~s. It encompasses three major events: 
a) The formation of the BH , \textcolor{black}{observation of the first GeV photon at $t_{\rm rf}=1.9$~s }, see details in \citet{2019ApJ...886...82R}.
b) \textcolor{black}{An increase of the 0.1-100 GeV luminosity following a power-law of $L_{\rm GeV} = 8.7 \times 10^{50}~t^{+(1.77\pm 0.28)}$~erg~s$^{-1}$}.
c) The \textcolor{black}{energetically dominant} UPE phase \textcolor{black}{observed by \textit{Fermi}-GBM} in the $10$~keV--$10$~MeV band, \textcolor{black}{occurring in the entire interval t$_{\rm rf}=$1.9s-3.99~s fulfilling a hierarchical structure} signed by a spectrum composed of a thermal emission and a cutoff power-law component (CPL$+$BB)\textcolor{black}{; see section~\ref{sec:heir}}. 

Episode 3, the ``cavity'', starts at $~t_{\rm rf}=11$~s and ends at $t_{\rm rf}=20$~s. The presence of a ``cavity'' in GRB 190114C, carved out in the SN ejecta by the BH formation, has been confirmed in \citet{2019ApJ...883..191R}. 

\textcolor{black}{ The GeV luminosity following the UPE phase} is best fitted by the \textcolor{black}{\emph{decreasing}} power-law of $ L_{\rm GeV} = \left(7.75\pm 0.44\right) \times 10^{52}~t^{-(1.2\pm 0.04)}$~erg~s$^{-1}$, with an isotropic energy of $E_{\rm GeV}= (1.8 \pm 1.3)\times 10^{53}$~erg. The spectrum of \textit{Fermi}-LAT in the $0.1$--$100$~GeV energy band, after the UPE phase, is best fitted by a power-law \citep{2019arXiv190107505W}; see Fig.~\ref{fig:data} and \citet{2021A&A...649A..75M} for more details.

All these results have been presented in \citet {2019arXiv190404162R}, \citet{2019ApJ...883..191R}, \citet{2019ApJ...886...82R} and \citet{2021A&A...649A..75M}.

\textcolor{black}{\section{The time-resolved spectral analysis, the hierarchical structure and the MeV luminosity of the UPE phase}\label{sec:heir}}

Following the spectral analysis performed over the UPE phase from $t_{\rm rf}=1.9$~s to $t_{\rm rf}=3.9$~s (first iteration), \textcolor{black}{we perform the} spectral analysis \textcolor{black}{over} the 1 second intervals ($\Delta t_{\rm rf}=1$~s), namely [$1.9$s--$2.9$s] and [$2.9$s--$3.9$s](second iteration).

Each half intervals are \textcolor{black}{further} divided in half (third iteration), i.e., $\Delta t_{\rm rf}=0.5$~s: [$1.9$s--$2.40$s], [$2.40$s--$2.9$s], [$2.9$s--$3.4$s] and [$3.4$s--$3.9$s] and the \textcolor{black}{corresponding} spectral analysis is performed over each interval.

We \textcolor{black}{further} divide the UPE into $8$ intervals of $\Delta t_{\rm rf}= 0.25$~s ( fourth iteration):  [$1.9$s--$2.15$s],[$2.15$s--$2.40$s], [$2.40$s--$2.65$s], [$2.65$s--$2.9$s], [$2.9$s--$3.15$s], [$3.15$s--$3.4$s], [$3.4$s--$3.65$s] and [$3.65$s--$3.9$s], and \textcolor{black}{perform} the same spectral analysis  over each interval. 

\textcolor{black}{We continue until} the final iteration (fifth iteration), \textcolor{black}{where the adequate signal-to-noise ratio S/N is fulfilled.} The UPE is divided into $16$ time intervals of $\Delta t_{\rm rf}= 0.125$~s: [$1.896$s--$2.019$s],[$2.019$s--$2.142$s],[$2.142$s--$2.265$s], [$2.265$s--$2.388$s], [$2.388$s--$2.511$s], [$2.511$s--$2.633$s], [$2.633$s--$2.756$s], [$2.756$s--$2.87$s], [$2.879$s--$3.002$s], [$3.002$s--$3.125$s], [$3.125$s--$3.248$s], [$3.248$s--$3.371$s], [$3.371$s--$3.494$s], [$3.494$s--$3.617$s], [$3.617$s--$3.739$s] and [$3.739$s--$3.862$s] and perform the spectral analysis. After dividing into subintervals of 0.125~s one extra time interval of [$3.862$s--$3.985$s] \textcolor{black}{has been added.} 

\begin{figure*}
\centering
\includegraphics[angle=90, width=0.84\hsize]{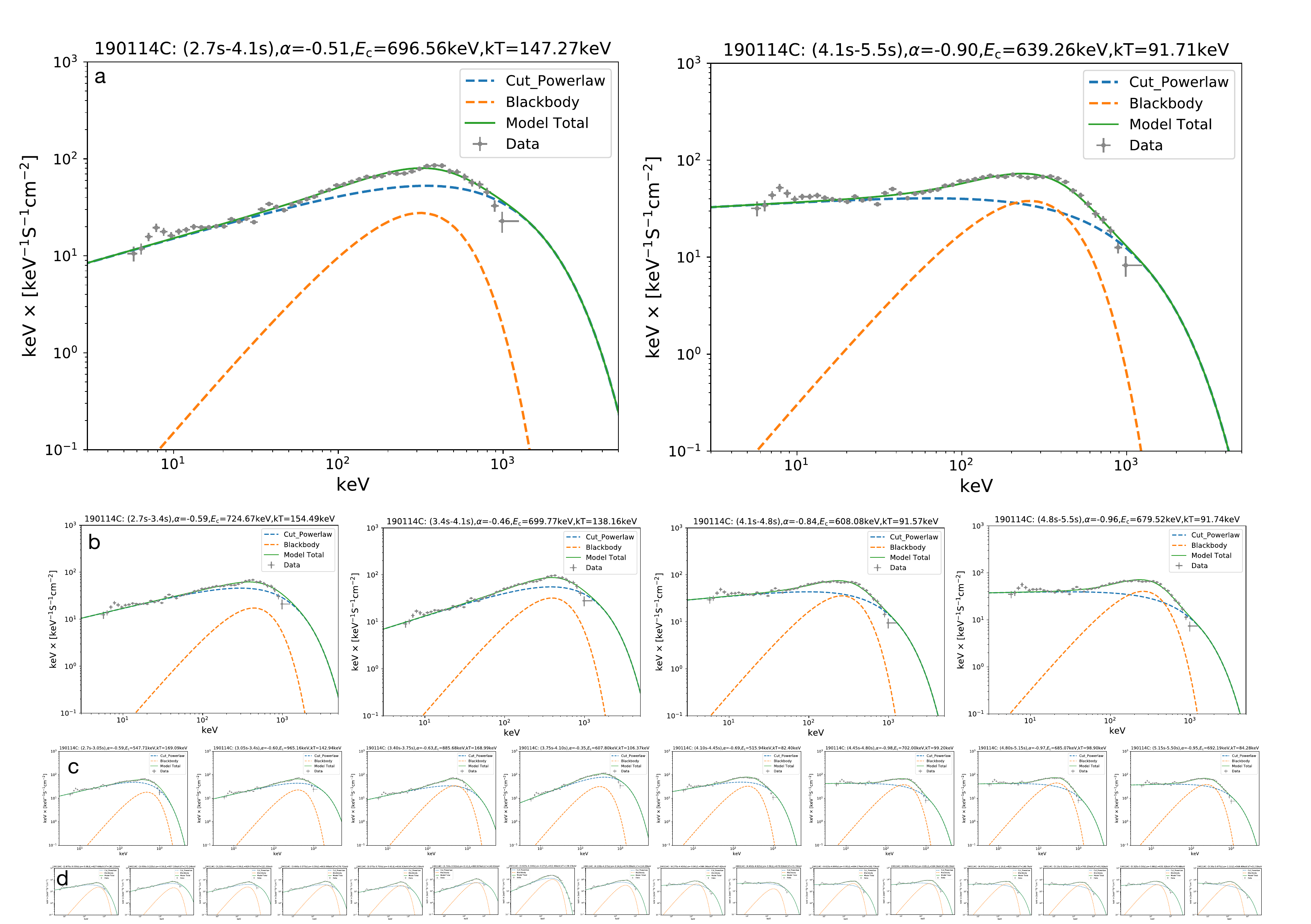}
\caption{Time-resolved spectral analysis of UPE phase GRB 190114C:  from $t=2.7$~s ($t_{\rm rf}=1.9$~s) to $t=5.5$~s ($t_{\rm rf}=3.9$~s). For the \textcolor{black}{second iteration}; (a) the time interval is divided into two parts, four parts for the \textcolor{black}{ third iteration}; (b), eight parts for the  \textcolor{black}{ fourth iteration}; (c), and sixteen parts  for the \textcolor{black}{fifth iteration}; (d), respectively. The spectral fitting \textcolor{black}{parameters for each iteration} are reported in Table \ref{tab:table}. \textcolor{black}{Plots are taken from \citet{2019arXiv190404162R} } with permission of authors.}\label{alltogether}
\end{figure*}

 The spectral fitting of a cutoff power law plus black body (CPL+BB) is confirmed in each time interval and for each iterative process; see Table~\ref{tab:table}, Fig.~\ref{alltogether}, \citet[][for more details]{2019arXiv190404162R}.

\begin{table*}[ht!]
\small\addtolength{\tabcolsep}{-0.5pt}
\caption{\textcolor{black}{The parameters of the time-resolved spectral  fits of the UPE phase of GRB 190114C, performed from $\Delta t_{\rm rf}= 2$~s down to subintervals of $\Delta t_{\rm rf}= 0.125$~s. The UPE phase extends}  from $t=2.7$~s ($t_{\rm rf}=1.9$~s) to $t=5.5$~s ($t_{\rm rf}=3.99$~s). \textcolor{black}{Column 1; represents the time intervals in the observer's frame (Obs), column 2; the time intervals in the rest-frame (rf), column 3; the statistical significance ($S$) for each time interval}, \textcolor{black}{ column 4; the power-law index of the cutoff power-law (CPL) component, column 5; the rest-frame cutoff energy, column 6; the rest-frame black body (BB) temperature, column 7; the Akaike Information Criterion/Bayesian Information Criterion (AIC/BIC), column 8; the BB flux ($F_{\rm BB}$), column 9; the CPL+BB or total flux ($F_{\rm tot}$), column 10; the ratio of the BB flux to the total flux, $F_{\rm BB}/F_{\rm tot}$ and, finally column 11; the isotropic energy in each time interval. As it can be seen from column 10, the} $F_{\rm BB}/F_{\rm tot}$ remains almost constant in each \textcolor{black}{iteration}. 
The AIC (\citealt{1974ITAC...19..716A}) and the BIC (\citealt{schwarz1978estimating}) \textcolor{black} {methods were used} to select non-nested and nested models, respectively \textcolor{black}{\citep[see][for more information about these methods]{Li2019,Li2019a,Li2019b,Li2020,Li2020b,Li2021}}.  \textcolor{black}{Table is taken from \citet{2019arXiv190404162R} with permission of authors.} \label{tab:table}}             
\label{tab:160509A}
\centering                         
\begin{tabular}{c|c|c|c|c|c|c|c|c|c|c}       
\hline\hline                  
$t_{1}$$\sim$$t_{2}$&$t_{rf,1}$$\sim$$t_{rf,2}$&{$S$}&$\alpha$&$E_{\rm c}$&$kT$&$\Delta$DIC&$F_{\rm BB}$&$F_{\rm tot}$&$F_{\rm ratio}$&$E_{\rm tot}$\\
\hline
(s)&(s)&&&(keV)&(keV)&&(10$^{-6}$)&(10$^{-6}$)&&(erg)\\
Obs&Rest-frame&&&&&&(erg~cm$^{-2}$~s$^{-1}$)&(erg~cm$^{-2}$~s$^{-1}$)\\ 
\hline                        
2.700$\sim$5.500&1.896$\sim$3.862&\textcolor{black}{418.62}&-0.71$^{+0.02}_{-0.02}$&717.6$^{+25.4}_{-25.4}$&159.0$^{+3.6}_{-3.6}$&-3344/6697/6719&22.49$^{+3.21}_{-2.65}$&111.10$^{+11.60}_{-10.40}$&0.20&1.50e+53\\
\hline
2.700$\sim$4.100&1.896$\sim$2.879&\textcolor{black}{296.60}&-0.51$^{+0.02}_{-0.02}$&696.6$^{+31.9}_{-32.4}$&209.7$^{+9.3}_{-9.1}$&-2675/5360/5381&24.67$^{+6.93}_{-5.35}$&142.50$^{+23.90}_{-21.00}$&0.17&9.64e+52\\
4.100$\sim$5.500&2.879$\sim$3.862&\textcolor{black}{318.07}&-0.90$^{+0.02}_{-0.02}$&639.3$^{+31.9}_{-31.6}$&130.6$^{+2.5}_{-2.5}$&-2529/5069/5090&25.55$^{+2.97}_{-2.75}$&80.98$^{+9.68}_{-8.07}$&0.32&5.48e+52\\
\hline
2.700$\sim$3.400&1.896$\sim$2.388&\textcolor{black}{204.30}&-0.59$^{+0.03}_{-0.03}$&724.7$^{+44.5}_{-45.5}$&220.0$^{+17.1}_{-17.2}$&-1882/3774/3796&18.55$^{+9.42}_{-7.40}$&123.90$^{+29.20}_{-22.30}$&0.15&4.19e+52\\
3.400$\sim$4.100&2.388$\sim$2.879&\textcolor{black}{225.88}&-0.46$^{+0.04}_{-0.04}$&699.8$^{+47.8}_{-48.3}$&196.7$^{+8.9}_{-8.7}$&-2032/4074/4095&31.78$^{+9.60}_{-7.31}$&161.40$^{+47.10}_{-32.40}$&0.20&5.46e+52\\
4.100$\sim$4.800&2.879$\sim$3.371&\textcolor{black}{233.97}&-0.84$^{+0.03}_{-0.03}$&608.1$^{+42.1}_{-42.2}$&130.4$^{+3.7}_{-3.9}$&-1880/3770/3792&23.94$^{+4.20}_{-4.22}$&85.37$^{+14.83}_{-12.27}$&0.28&2.89e+52\\
4.800$\sim$5.500&3.371$\sim$3.862&\textcolor{black}{227.90}&-0.96$^{+0.03}_{-0.03}$&679.5$^{+49.1}_{-48.7}$&130.6$^{+3.1}_{-3.2}$&-1809/3628/3649&27.18$^{+4.01}_{-3.73}$&78.20$^{+11.40}_{-9.66}$&0.35&2.65e+52\\
\hline
2.700$\sim$3.050&1.896$\sim$2.142&\textcolor{black}{148.59}&-0.59$^{+0.03}_{-0.03}$&547.7$^{+44.2}_{-44.9}$&240.8$^{+29.2}_{-29.1}$&-1187/2384/2406&19.67$^{+17.96}_{-8.88}$&103.20$^{+30.60}_{-20.28}$&0.19&1.75e+52\\
3.050$\sim$3.400&2.142$\sim$2.388&\textcolor{black}{145.04}&-0.60$^{+0.02}_{-0.02}$&965.2$^{+28.5}_{-30.1}$&203.5$^{+14.8}_{-14.8}$&-1320/2650/2671&22.87$^{+8.88}_{-7.23}$&152.00$^{+24.00}_{-21.00}$&0.15&2.57e+52\\
3.400$\sim$3.750&2.388$\sim$2.633&\textcolor{black}{134.60}&-0.63$^{+0.04}_{-0.04}$&885.7$^{+70.9}_{-70.1}$&240.6$^{+10.5}_{-10.6}$&-1224/2458/2480&41.02$^{+11.09}_{-7.91}$&129.10$^{+32.40}_{-23.40}$&0.32&2.18e+52\\
3.750$\sim$4.100&2.633$\sim$2.879&\textcolor{black}{187.77}&-0.35$^{+0.06}_{-0.05}$&607.8$^{+57.1}_{-60.1}$&151.5$^{+12.4}_{-14.2}$&-1428/2866/2887&23.92$^{+12.46}_{-10.40}$&192.00$^{+101.70}_{-60.30}$&0.12&3.25e+52\\
4.100$\sim$4.450&2.879$\sim$3.125&\textcolor{black}{171.81}&-0.69$^{+0.04}_{-0.04}$&515.9$^{+43.6}_{-43.6}$&117.3$^{+5.0}_{-5.0}$&-1271/2552/2573&19.19$^{+4.89}_{-4.40}$&92.71$^{+27.69}_{-22.43}$&0.21&1.57e+52\\
4.450$\sim$4.800&3.125$\sim$3.371&\textcolor{black}{230.14}&-0.98$^{+0.04}_{-0.04}$&702.0$^{+78.1}_{-78.2}$&141.3$^{+5.8}_{-5.8}$&-1254/2518/2539&26.76$^{+6.41}_{-5.47}$&80.73$^{+17.95}_{-14.95}$&0.33&1.37e+52\\
4.800$\sim$5.150&3.371$\sim$3.617&\textcolor{black}{166.30}&-0.97$^{+0.04}_{-0.04}$&685.1$^{+69.4}_{-68.6}$&140.8$^{+4.6}_{-4.6}$&-1218/2447/2468&31.83$^{+6.85}_{-4.98}$&82.51$^{+15.62}_{-12.33}$&0.39&1.40e+52\\
5.150$\sim$5.500&3.617$\sim$3.862&\textcolor{black}{161.51}&-0.95$^{+0.04}_{-0.04}$&692.2$^{+79.1}_{-77.7}$&120.0$^{+4.0}_{-4.0}$&-1203/2416/2438&23.19$^{+5.38}_{-3.81}$&73.57$^{+18.69}_{-12.93}$&0.32&1.24e+52\\
\hline
2.700$\sim$2.875&1.896$\sim$2.019&\textcolor{black}{117.09}&-0.58$^{+0.05}_{-0.05}$&470.5$^{+74.4}_{-83.7}$&261.5$^{+29.0}_{-27.9}$&-640/1291/1311&33.68$^{+20.39}_{-14.33}$&112.30$^{+28.37}_{-25.73}$&0.30&9.50e+51\\
2.875$\sim$3.050&2.019$\sim$2.142&\textcolor{black}{94.40}&-0.68$^{+0.04}_{-0.05}$&627.6$^{+87.0}_{-91.5}$&258.0$^{+30.1}_{-28.7}$&-664/1337/1359&28.45$^{+20.42}_{-12.51}$&98.14$^{+33.56}_{-26.44}$&0.29&8.30e+51\\
3.050$\sim$3.225&2.142$\sim$2.265&\textcolor{black}{106.62}&-0.59$^{+0.03}_{-0.03}$&957.1$^{+34.1}_{-34.9}$&245.3$^{+21.5}_{-21.0}$&-768/1547/1568&25.71$^{+13.87}_{-9.03}$&169.30$^{+38.20}_{-31.60}$&0.15&1.43e+52\\
3.225$\sim$3.400&2.265$\sim$2.388&\textcolor{black}{100.40}&-0.73$^{+0.06}_{-0.06}$&1275.9$^{+208.9}_{-215.4}$&208.6$^{+9.1}_{-9.2}$&-669/1349/1369&36.78$^{+9.54}_{-8.93}$&144.90$^{+33.02}_{-27.63}$&0.25&1.23e+52\\
3.400$\sim$3.575&2.388$\sim$2.511&\textcolor{black}{98.23}&-0.59$^{+0.05}_{-0.05}$&804.0$^{+86.7}_{-82.3}$&255.9$^{+17.4}_{-17.4}$&-702/1414/1436&42.19$^{+19.41}_{-13.59}$&139.30$^{+48.30}_{-35.60}$&0.30&1.18e+52\\
3.575$\sim$3.750&2.511$\sim$2.633&\textcolor{black}{93.84}&-0.65$^{+0.04}_{-0.04}$&916.3$^{+64.6}_{-67.7}$&229.3$^{+13.6}_{-13.5}$&-730/1471/1492&39.25$^{+11.97}_{-10.71}$&119.50$^{+32.90}_{-25.45}$&0.33&1.01e+52\\
3.750$\sim$3.925&2.633$\sim$2.756&\textcolor{black}{126.63}&-0.51$^{+0.02}_{-0.02}$&960.9$^{+30.9}_{-31.4}$&204.6$^{+9.9}_{-10.0}$&-808/1627/1648&57.70$^{+15.81}_{-12.25}$&221.10$^{+35.60}_{-31.50}$&0.26&1.87e+52\\
3.925$\sim$4.100&2.756$\sim$2.879&\textcolor{black}{141.61}&-0.27$^{+0.06}_{-0.06}$&412.7$^{+12.2}_{-11.9}$&196.8$^{+14.0}_{-16.1}$&-729/1468/1488&32.20$^{+19.05}_{-18.86}$&176.50$^{+12.91}_{-11.21}$&0.18&1.49e+52\\
4.100$\sim$4.275&2.879$\sim$3.002&\textcolor{black}{122.91}&-0.54$^{+0.06}_{-0.06}$&474.1$^{+45.5}_{-46.2}$&162.6$^{+14.9}_{-14.8}$&-758/1526/1547&24.26$^{+17.09}_{-10.09}$&116.10$^{+52.40}_{-35.12}$&0.21&9.82e+51\\
4.275$\sim$4.450&3.002$\sim$3.125&\textcolor{black}{122.62}&-0.64$^{+0.08}_{-0.08}$&365.0$^{+44.9}_{-48.5}$&107.5$^{+15.7}_{-12.6}$&-675/1360/1380&9.04$^{+9.47}_{-5.69}$&72.20$^{+19.06}_{-14.95}$&0.13&6.11e+51\\
4.450$\sim$4.625&3.125$\sim$3.248&\textcolor{black}{111.94}&-1.04$^{+0.05}_{-0.05}$&640.0$^{+108.7}_{-106.1}$&161.0$^{+11.1}_{-10.8}$&-640/1290/1310&22.34$^{+9.36}_{-6.65}$&68.54$^{+11.70}_{-11.21}$&0.33&5.80e+51\\
4.625$\sim$4.800&3.248$\sim$3.371&\textcolor{black}{123.33}&-0.95$^{+0.05}_{-0.05}$&694.2$^{+96.8}_{-94.2}$&146.3$^{+6.7}_{-6.6}$&-734/1477/1499&35.59$^{+9.47}_{-8.00}$&89.91$^{+27.59}_{-18.82}$&0.40&7.60e+51\\
4.800$\sim$4.975&3.371$\sim$3.494&\textcolor{black}{129.65}&-0.85$^{+0.05}_{-0.05}$&564.5$^{+68.9}_{-71.9}$&135.3$^{+7.5}_{-7.6}$&-744/1498/1519&30.78$^{+11.12}_{-8.55}$&96.58$^{+31.02}_{-23.68}$&0.32&8.17e+51\\
4.975$\sim$5.150&3.494$\sim$3.617&\textcolor{black}{107.36}&-1.10$^{+0.04}_{-0.04}$&820.5$^{+115.0}_{-111.2}$&149.7$^{+5.9}_{-5.8}$&-683/1376/1398&32.76$^{+6.98}_{-5.92}$&71.57$^{+16.74}_{-11.99}$&0.46&6.05e+51\\
5.150$\sim$5.325&3.617$\sim$3.739&\textcolor{black}{108.96}&-1.04$^{+0.05}_{-0.05}$&765.2$^{+119.0}_{-115.8}$&130.9$^{+5.8}_{-5.8}$&-697/1404/1426&26.14$^{+7.02}_{-5.96}$&66.70$^{+20.48}_{-14.17}$&0.39&5.64e+51\\
5.325$\sim$5.500&3.739$\sim$3.862&\textcolor{black}{121.57}&-0.88$^{+0.06}_{-0.06}$&635.3$^{+88.7}_{-92.0}$&108.9$^{+5.3}_{-5.4}$&-736/1483/1504&20.90$^{+6.51}_{-5.15}$&79.48$^{+28.02}_{-21.03}$&0.26&6.72e+51\\
\hline                                   
\end{tabular}
\end{table*}
\begin{figure*}
\centering
\textcolor{black}{[a]}\includegraphics[width=0.47\hsize,clip]{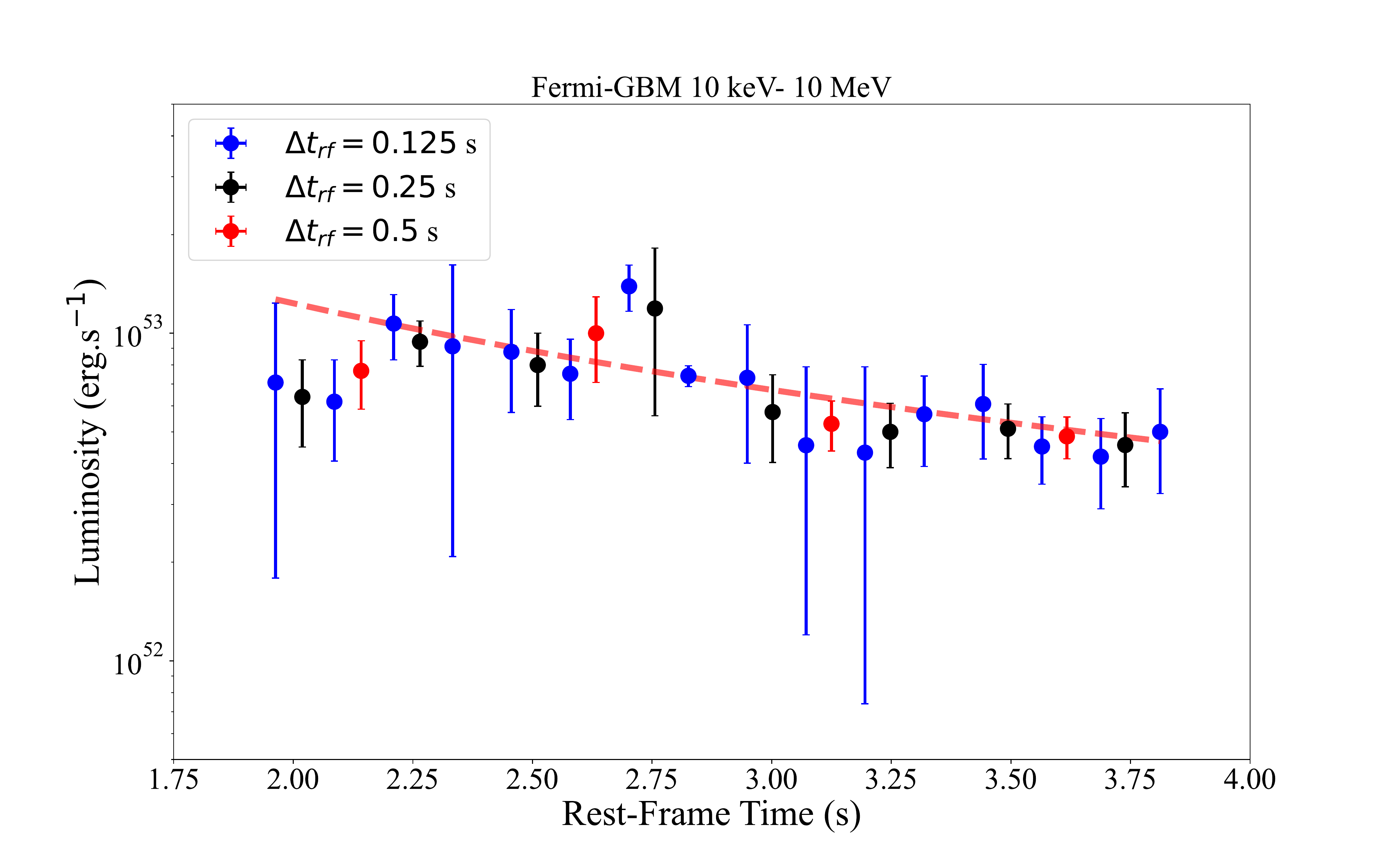}
\textcolor{black}{[b]}\includegraphics[width=0.47\hsize,clip]{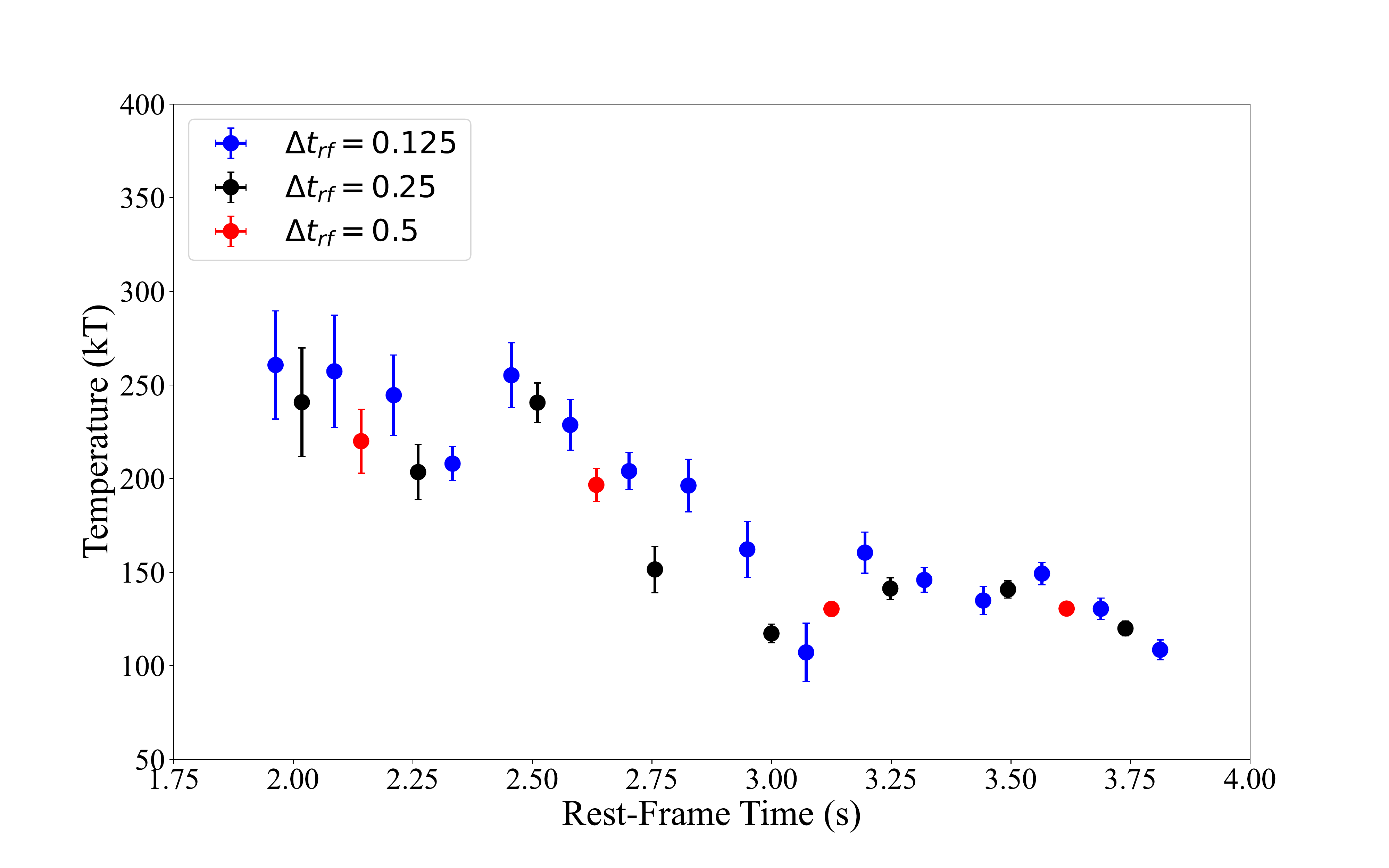}
\caption{Luminosity \textcolor{black}{[a] and rest-frame temperature [b] during} the UPE observed by \textit{Fermi}-GBM, \textcolor{black}{obtained from analyses with $\Delta t=0.125$~s (blue circles), $\Delta t=0.25$~s (black circles) and $\Delta t=0.5$~s (red circles) time resolutions reported in Table~\ref{tab:table}}. The luminosity is best fitted by a power-law of amplitude $(3.5 \pm 1.1) \times 10^{53}$~erg~s$^{-1}$ and power-law index $-1.50 \pm 0.30$. \textcolor{black}{The best fit of luminosity, obtained from $\Delta t_{\rm rf}= 0.125$~s time-resolved analysis, is in principle independent of the resolution of data analysis and is fulfilled in all iterative sequences.}}
\label{fig:lumupe}
\end{figure*}
From a time-resolved analysis of the UPE phase, performed down to the fifth iteration, a hierarchical structure is obtained. It reveals a common spectral feature for each subinterval characterized by the CPL$+$BB best-fit model with a rest-frame temperature of $kT= 100\sim 300$~keV and the ratio of blackbody flux \textcolor{black}{($F_{\rm BB}$)} to the total flux \textcolor{black}{($F_{\rm tot}$)} of:
\begin{equation}\label{eq:ratio} 
0.1\lesssim \frac{F_{\rm BB}}{F_{\rm tot}} \lesssim 0.5.
\end{equation}
see Table~\ref{tab:table}, \textcolor{black}{ Fig.~\ref{alltogether} and} \citet[][for more information]{2019arXiv190404162R}

During the UPE phase the MeV luminosity is best fitted by
\begin{equation}
 L_{\rm MeV} =A_{\rm MeV}~t^{-\alpha_{\rm MeV}}, 
 \label{Mevluminosity}
\end{equation}
with slope $\alpha_{\rm MeV}=1.5\pm 0.3$, and amplitude $A_{\rm MeV}=(3.5\pm 1.1) \times 10^{53}$ erg~s$^{-1}$. \textcolor{black}{This best fit is obtained from $\Delta t_{\rm rf}= 0.125$~s time-resolved analysis and is fulfilled in all iterative sequences}; see \textcolor{black}{Table \ref{tab:160509A}} and Figs.~\ref{fig:data} and \ref{fig:lumupe}.

The existence of the thermal \textcolor{black}{and the cutoff power-law}  components in the \textcolor{black}{spectra} of the UPE phase \textcolor{black}{have been identified as the characteristic signature of $e^+~e^-$ pair creation in presence of baryons (PEMB pulse) originating from the vacuum polarization process \cite{1999A&AS..138..511R,1999A&A...350..334R, 2000A&A...359..855R,2010PhR...487....1R}}; see Sec.~\ref{sec:vac}.

\section{The properties of \emph{inner engine} }\label{sec:5}

\textcolor{black}{The discovery of the \emph{inner engine} obtained by} incorporating the Papapetrou-Wald solution \citep{1966AIHPA...4...83P,1974PhRvD..10.1680W, 2019ApJ...886...82R, 2020EPJC...80..300R,2021A&A...649A..75M}, around the newborn Kerr BH in the BdHNe I, \textcolor{black}{in presence of the low density}  plasma  of the cavity, see \citep{2019ApJ...883..191R}, \textcolor{black}{was operative in the classical electrodynamics with $|{\bf E}|<E_c$ process and leading to the generation by synchrotron radiation of the GeV emission in GRB 130427A and GRB190114C. We also apply here this \textit{inner engine} in the $|{\bf E}|>E_c$ regime to describe the quantum electrodynamics process.}

Wald's work is based on the Papapetrou discovery \cite{1966AIHPA...4...83P} that Killing vectors are vector potential solutions of sourceless Maxwell equations in vacuum spacetimes in the test field approximation (i.e. no metric backreaction). \textcolor{black}{A} linear combination of these two Killing vector solutions led Wald to the solution for a rotating BH immersed in a uniform magnetic field \textcolor{black}{ $B_0$, aligned with the rotation axis of the Kerr BH} at infinity.

The electromagnetic field of the \emph{inner engine}  in the Carter's orthonormal tetrad is:
\begin{eqnarray}
   E_{\hat{r}} &=& \frac{\hat{a} B_0}{\Sigma} \left[r\sin^2\theta-\frac{\hat{M} \left(\cos ^2\theta+1\right) \left(r^2-\hat{a}^2 \cos ^2\theta \right)}{\Sigma}\right],\\
    E_{\hat{\theta}}&=&\frac{\hat{a} B_0}{\Sigma}\sin\theta \cos\theta \sqrt{\Delta},\\ 
  B_{\hat{r}}&=&-\frac{B_0 \cos\theta}{\Sigma} \left[-\frac{2 \hat{a}^2 \hat{M} r \left(\cos ^2\theta+1\right)}{\Sigma }+\hat{a}^2+r^2\right],\\
  B_{\hat{\theta}}&=&  \frac{B_0 r}{\Sigma}\sin\theta \sqrt{\Delta },
 \end{eqnarray}
where $\Sigma=r^2+\hat{a}^2\cos^2\theta$, $\Delta=r^2-2 \hat{M} r+\hat{a}^2$, $\hat{M}= G M/c^2$, $\hat{a}=a/c=J/(M\,c)$, being $M$ and $J$ the mass and angular momentum of the Kerr BH. The (outer) event horizon is located at $r_+=(\hat{M}+\sqrt{\hat{M}^2-\hat{a}^2})$. 
 
The electromagnetic field in the polar direction $\theta=0$ and at small angles from it is well approximated by \citep{2019ApJ...886...82R, 2020EPJC...80..300R}:
\begin{eqnarray}
   E_{\hat{r}} &=& -\frac{2 B_0 J\,G}{c^3} \frac{ \left(r^2-\hat{a}^2 \right)}{\left(r^2+\hat{a}^2 \right)^2} \label{eq:ER} \\ 
    E_{\hat{\theta}}&=&0 \\
  B_{\hat{r}}&=&\frac{B_0  \left(-\frac{4\,G\, J^2 r}{M\left(r^2+\hat{a}^2 \right) }+a^2+r^2\right)}{\left(r^2+\hat{a}^2 \right)}\\
  B_{\hat{\theta}}&=& 0.
 \end{eqnarray}
Equation~(\ref{eq:ER}) is the same as the radial electric field of the Kerr-Newman metric in the same tetrad  \textcolor{black}{just substituting to the charge $Q$ of the Kerr-Newman solution the effective charge $Q_{\rm eff}$, given by Eq.~(\ref{eq:EFCH}}), see e.g.  \cite{1978pans.proc.....G}.
Therefore, up to linear order in $\theta$ and in the dimensionless BH spin parameter $\alpha \equiv \hat{a}/(G M/c^2)$, the electric field can be written as
\begin{equation}\label{eq:ER2}
   E_{\hat{r}} = -\frac{2 B_0 J\,G}{c^3} \frac{ \left(r^2-\hat{a}^2 \right)}{\left(r^2+\hat{a}^2 \right)^2}\approx -\frac{1}{2}\alpha B_0\frac{r_+^2}{r^2},
\end{equation}
which for spin values $\alpha\lesssim 0.7$, \textcolor{black}{the available electrostatic energy} is well approximated by
\begin{equation}
\mathcal{E} \approx \frac{\left( 2 B_0 J G/c^3\right)^2}{2\,r_+}=\textcolor{black}{\frac{Q_{\rm eff}^2}{2\,r_+}}=1.25 \times 10^{43}\frac{\beta^2 \alpha^2 \mu^3}{1+\sqrt{1-\alpha^2}}  \textrm{ erg},
\label{approx}
\end{equation}
where we have normalized the mass and magnetic field strength by, respectively, $\mu = M/M_\odot$ and $\beta=B_0/B_c$, being 
\begin{equation}
    B_c=E_c=\frac{m_e^2 c^3}{e\hbar} \approx 4.41\times 10^{13}\,\rm G,
\end{equation}
the critical field for vacuum polarization\textcolor{black}{; see \citet{2019ApJ...886...82R} for more details}. 

\textcolor{black}{The values $\mu$ and $\beta$ in this general equations will be determined as a function of astrophysical process operative in the \textit{inner engine} in GRB 190114C. The corresponding description of the overcritical and the undercritical fields are represented in the next sections.}

We adopt that the magnetic field and BH spin are parallel, therefore along the symmetry axis direction electrons in the surrounding ionized medium are repelled, while protons are pulled into the BH \citep[see][for additional details]{2019ApJ...886...82R}.

\section{The mass and spin of the BH of GRB 190114C}\label{sec:mass-spin}

We here recall the self-consistent solution \textcolor{black}{following the UPE phase}, well tested in the case of GRB 130427A \citep{2019ApJ...886...82R} and GRB 190114C \citep{2021A&A...649A..75M} which fulfills three conditions:
1) The GeV energetics observed by the Fermi-LAT is paid by the extractable energy of the BH, i.e.: $E_{\rm GeV}= E_{\rm extr}$. 2) The magnetic $e^+~e^-$ pair production (MPP) process does not occur around the BH, therefore the GeV photons fulfill the transparency condition. 3) The timescale of the synchrotron radiation determines the time scale of observed GeV radiation.

Having these conditions, and assuming the minimum energy budget requirement;, the \emph{inner engine}  parameters at $t>t_{\rm rf}=3.99~$s, \textcolor{black}{i.e., after the UPE phase,} are: magnetic field strength $B_0\approx 3.9\times 10^{10}$~G, spin and BH mass, respectively, $\alpha = 0.41$ and $M=4.45~M_\odot$. The corresponding BH irreducible mass is $M_{\rm irr}=4.35~M_\odot$ see \citet{2021A&A...649A..75M} \textcolor{black}{for more details}. 

\section{Determination of the mass and spin of the BH during the UPE phase}\label{sec:massupe} 

We have \textcolor{black}{obtained in the previous section}  at $t_{\rm rf}=3.99$~s the values of mass and spin parameters of the BH and the magnetic field: $M=4.45 M_\odot$, $\alpha = 0.41$, and $B_0=3.9 \times 10^{10}$~G, respectively.  \textcolor{black}{We now turn to the determination of the mass and spin of the BH during the UPE phase.}
The mass-energy formula of the Kerr BH (\citealp{1970PhRvL..25.1596C,1971PhRvD...4.3552C,1971PhRvL..26.1344H}; see also ch.33 in \citep{1973grav.book.....M}) is given by:
\begin{equation}
\label{aone}
M^2 = \frac{c^2 J^2}{4 G^2 M^2_{\rm irr}}+M_{\rm irr}^2.
\end{equation}

We require that the energetics of the MeV radiation be explained by the extractable rotational energy of the Kerr BH, i.e., 
\begin{equation}
\label{Eextr1}
E_{\rm MeV} = E_{\rm extr}= (M-M_{\rm irr})c^2.
\end{equation}

Therefore, the extractable energy is given by:
\begin{equation}
\label{Eextr}
E_{\rm extr}=(M-M_{\rm irr}) c^2=\left(1-\sqrt{\frac{1+\sqrt{1-\alpha^2}}{2}}\right)M c^2.
\end{equation}

 The time derivative of Eq.~(\ref{Eextr}) gives the luminosity
\begin{equation}
\label{sdown1}
{L_{\rm MeV}=-\frac{dE_{extr}}{dt}=-\frac{dM}{dt},}
\end{equation}
in which we assume that $M_{irr}$ is constant for BH during the energy emission process.   

From the luminosity of MeV radiation expressed in the rest-frame of the GRB, given by Eq.~(\ref{Mevluminosity}) in the time interval of the UPE phase (see also Fig.~\ref{fig:lumupe}), and from the values of the spin and of the mass of the BH at $t_{\rm rf}=3.99$~s,  we can now work backward by integrating Eq.~(\ref{sdown1}) and determine the BH mass and spin at the beginning of the UPE, when the BH is formed, namely at $t_{\rm rf}=1.9$~s. We obtain $M=4.53 M_\odot$ and $\alpha=0.54$, respectively. 

This assumption demands that all the luminosity of UPE phase originates \textcolor{black}{from} the rotational energy of the BH. This point is going to be justified in the next sections.

\section{Vacuum polarization, dyadoregion and the UPE phase} \label{sec:vac}

The UPE phase is characterized by an electric field $|{\bf E}|> E_c$ \citep{2010PhR...487....1R,2018ApJ...869..151R}. The problem of vacuum polarization due to the overcritical field has a vast literature which dates back to the concept of the \textcolor{black}{the dyadosphere \citep{1998astro.ph.11232R} and} dyadotorous in the Kerr-Newman geometry, developed in \citet{2009PhRvD..79l4002C}. \textcolor{black}{Dyado is from the Greek word ``duados'' for pair, indicating here the $e^+~e^-$ pairs.} The dyadotorus is the region where the vacuum polarization processes occur around a rotating charged BH, leading to the production of $e^+~e^-$ pairs; see also \citep{2010PhR...487....1R} for details. 
 
In order to evaluate this process in the present case, we adopt a description using the Kerr-Newman geometry for which an analytic formula for the energy contained in the dyadoregion has been derived in \citet{2009PhRvD..79l4002C}. We have checked numerically that the energy of the dyadoregion in the Kerr-Newman geometry (see Eq.~\ref{Eemxi} below), setting the BH charge as the effective charge of the Papapetrou-Wald solution, $2 B_0 J G/c^3$, is a good approximation of the one estimated numerically with the Papapetrou-Wald solution. We have verified that the quantitative difference is at most $30\%$, which implies that this approximation does not affect our conclusions.

We can now evaluate the energy of $e^+~e^-$ pairs generated in the Papapetrou-Wald solution using the Kerr-Newman analogy. We use the Carter orthonormal frame, in which the flat spacetime Schwinger framework can be locally applied and determine the dyadoregion energy \citep[see discussion in][]{2009PhRvD..79l4002C}:
\begin{eqnarray}
\label{Eemxi}
E_{(r_+,r_{\rm d})}
&=&\frac{(2 B_0 J G/c^3)^2}{4r_+}\left(1-\frac{r_+}{r_{\rm d}}\right)+\frac{(2 B_0 J G/c^3)^2}{4\hat{a}}\nonumber\\
&\times&\left[\left(1+\frac{\hat{a}^2}{r_+^2}\right)\arctan\left(\frac{\hat{a}}{r_+}\right)\right.\nonumber\\
&-&\left.\left(1+\frac{\hat{a}^2}{r_d^2}\right)\arctan\left(\frac{\hat{a}}{r_d}\right)\right],
\end{eqnarray}
where $r_d$ is the radius of the dyadoregion
\begin{equation}
\label{dyadosurf}
\left(\frac{r_d}{\hat{M}}\right)^2=\frac12\frac{\lambda}{\mu\epsilon}
-\alpha^2+\left(\frac14\frac{\lambda^2}{\mu^2\epsilon^2}
-2\frac{\lambda}{\mu\epsilon}\alpha^2\right)^{1/2}\,
\end{equation}
with $\epsilon=  E_c M_\odot G^{3/2}/c^4\approx 1.873 \times10^{-6}$, and 
\begin{equation}
\label{eq:lamda}
\lambda = \frac{2 B_0 J G/c^3}{\sqrt{G} M} = \textcolor{black}{\frac{Q_{\rm eff}}{\sqrt{G} M}},    
\end{equation}
is the effective charge-to-mass ratio.

The characteristic width of the \textit{dyadoregion}, \textcolor{black}{i.e. the region around the BH where the electric field overcritical} is
\begin{equation}
\label{eq:width}
 \Delta_{\rm d}(t)= r_d(t)-r_+(t).
\end{equation}
\begin{figure*}
\centering
[a]\includegraphics[width=8.0cm]{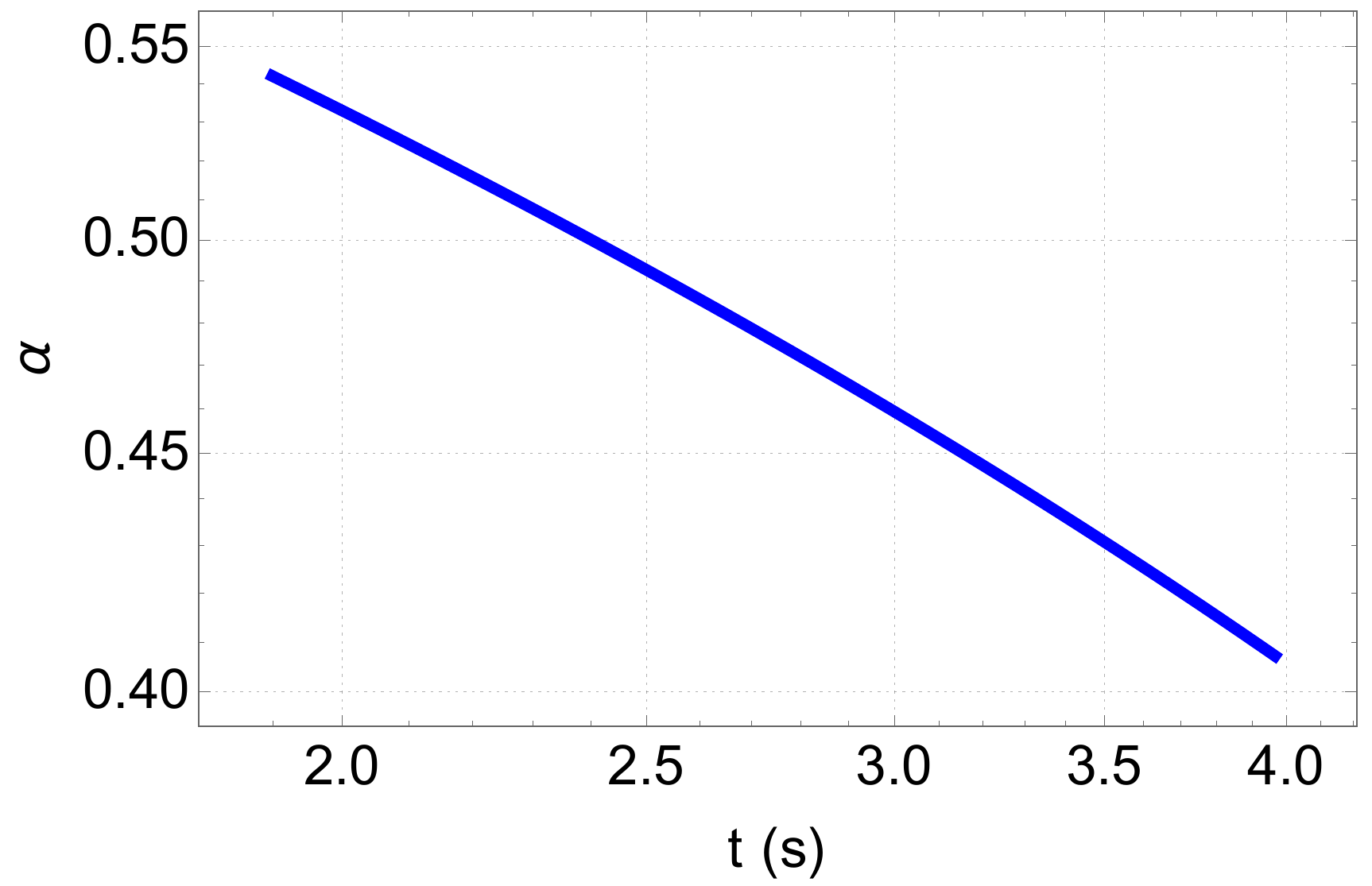}
[b]\includegraphics[width=8.0cm]{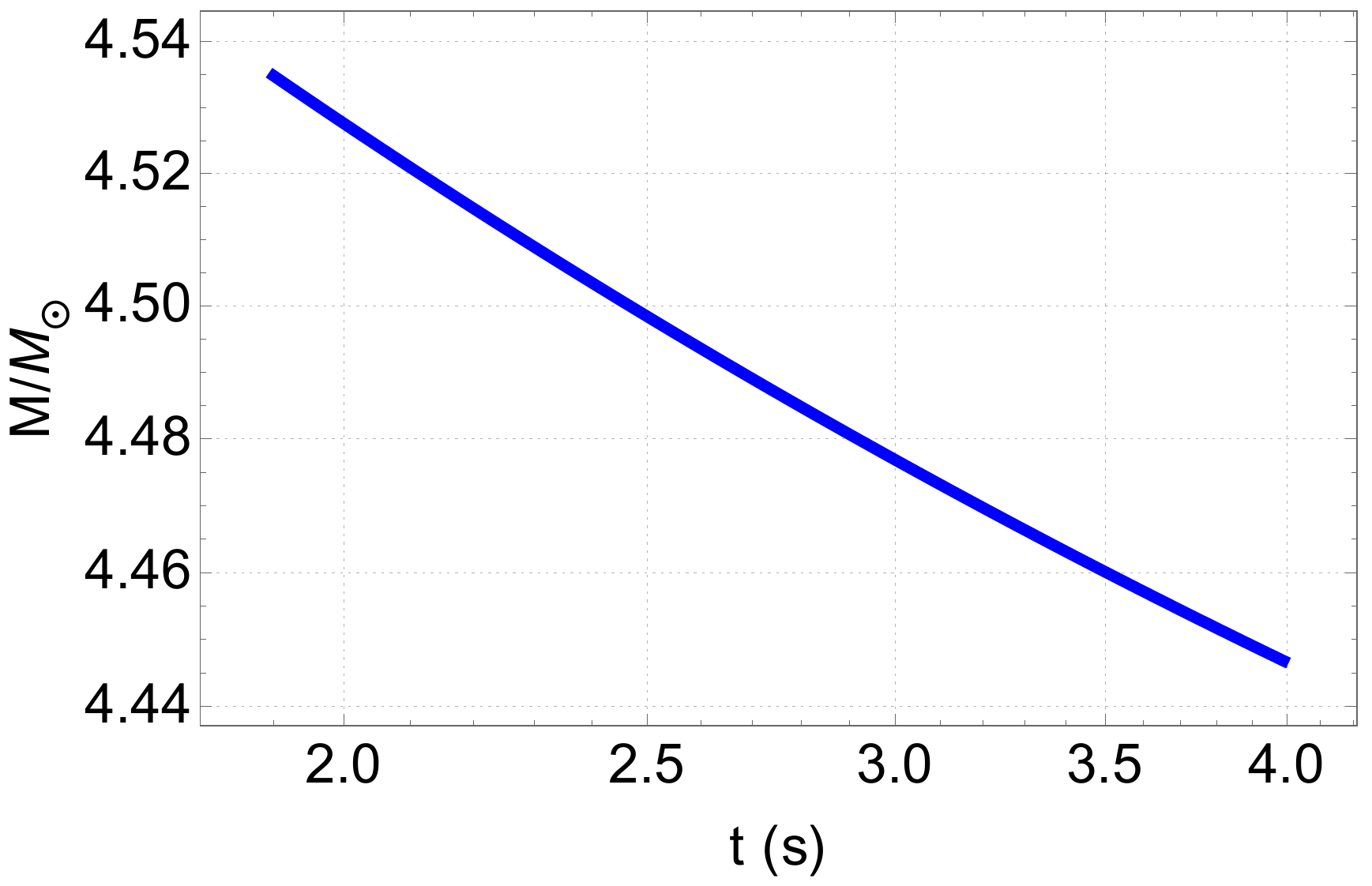}
\caption{ 
The decrease of the BH spin and mass, as a function of rest-frame time for GRB 190114C during the UPE phase, namely in the rest-frame time interval $t_{\rm rf}=1.9$--$3.99$~s. The values of spin and mass at the moment when BH formed are, respectively, $M=4.53 M_\odot$ and $\alpha=0.54$. At the moment when the UPE is over, i.e. at $t_{\rm rf}=3.99$~s, are: $\alpha=0.41$ and $M=4.45 M_{\odot}$.}
\label{massspinupe}
\end{figure*}
%

\section{Transparency condition in the UPE phase}\label{sec:12}

The existence of overcritical fields in the UPE phase and the consequent production of an $e^+~e^-~\gamma$ plasma, have been \textcolor{black}{addressed} in Sec.~\ref{sec:vac}.

In presence of an overcritical electric field around the BH, a sequence of events occur:

1) An optically thick $e^+~e^-~\gamma$ plasma of total energy $E_{e^+~e^-}^{\mathrm{tot}}=E_{\gamma, \rm iso}$ endowed with baryon load with a mass of $M_B$. The self-acceleration and expansion of such \textcolor{black}{\textit{PEMB pulses}} has been described in \citet{1999A&A...350..334R}. The dynamics of the \textcolor{black}{PEMB pulses} due to the effect of baryonic matter (the remnant of the collapsed object) has been considered in \cite{,2000A&A...359..855R}.  The thermalization of the pair plasma is achieved almost instantaneously ($\sim 10^{-13}$~s) and expands due to its self-acceleration up to ultrarelativistic velocities ($\Gamma \sim 100$ in the case of long GRBs; \citealp{2007PhRvL..99l5003A,2009PhRvD..79d3008A}).

2) The transparency of the $e^+~e^-~\gamma$ plasma. When the \textcolor{black}{PEMB pulses} expand with ultrarelativistic velocities, the $e^+~e^-~\gamma$ plasma becomes optically thin, a thermal radiation that has been called the Proper-GRB (P-GRB) is emitted \citep{1999A&A...350..334R,2000A&A...359..855R}. The P-GRB is characterized by the observed thermal component; see Sec.~\ref{sec:3} and \textcolor{black}{Sec.~\ref{sec:heir}}. The dynamics of the expanding plasma from the vicinity of the BH up to the transparency point is described by the plasma energy, $E_{e^+e^-}^{\mathrm{tot}}$ and the baryon load parameter, ${\cal B}=M_Bc^2/E_{e^+e^-}^{\mathrm{tot}}$\citep{1999A&A...350..334R,2000A&A...359..855R}.

The total P-GRB energy in the comoving frame of each impulsive process is
\begin{eqnarray}
\label{eq:energy1}
E^{com}_{\rm P-GRB}&=& \int a T^4_{com} dV_{com}, \nonumber\\
&=& a T^4_{com}V_{com}
\label{dyaE}
\end{eqnarray}
where $a$ is radiation constant, $T_{com}$ is the P-GRB temperature in the comoving frame and $V_{com}$ is the volume of the \textcolor{black}{PEMB pulses} in the comoving frame.

Dividing Eq.~(\ref{eq:energy1}) by the Doppler factor $\Gamma(1-v/c)$ at transparency, i.e. when the P-GRB is emitted, being $\Gamma$ and $v$ the Lorentz factor and speed of the \textcolor{black}{PEMB pulses}, and assuming head-on emission; namely $\cos\vartheta=1$ one can obtain:
\begin{equation}
\label{eq:energy2}
\frac{E^{com}_{\rm P-GRB}}{\Gamma(1-v/c)}= \frac{a T^4_{com}}{\Gamma(1-v/c)} V_{com}\, ,
\end{equation} 
where we have assumed head-on emission and therefore fixed $\cos\vartheta=1$ in the Doppler factor. 

Since: 
\begin{eqnarray}
T_{obs} &=& \frac{T_{com}}{\Gamma(1-v/c)}\, ,\nonumber\\
E^{obs} &=& \frac{E^{com}}{\Gamma(1-v/c)}\, ,\nonumber\\
V_{lab} &=& \frac{V_{com}}{\Gamma}\, ,
\end{eqnarray}
we have that:
\begin{eqnarray}
\label{eq:energy3}
E^{obs}_{\rm P-GRB} &=& a T^4_{obs} \Gamma^3(1-v/c)^3  \Gamma V_{lab}\nonumber \\
&=& a T^4_{obs} \Gamma^4(1-v/c)^3  4\pi R^2 \Delta_{lab},
\end{eqnarray}
where we have used the fact that $V_{lab}=4\pi R^2 \Delta_{lab}$, where $\Delta_{lab}$ is the thickness of the \textcolor{black}{PEMB pulses}, and $a=4\sigma/c$, being $\sigma$ the Stefan-Boltzmann constant.

Moreover, we know from the condition of transparency 
\begin{eqnarray}
\label{eq:transp1}
\tau &=&  \sigma_T (n_{e^+e^-}+\bar{Z} n_B)\Delta_{lab} \approx \sigma_T (\bar{Z} n_B) \Delta_{lab},\nonumber \\
& = & \sigma_T \frac{\bar{Z} M_B }{m_N 4 \pi R^2  \Delta_{lab}} \Delta_{lab}  = 1,
\end{eqnarray}
where $\sigma_T$ is the Thomson cross section, $\bar{Z}$ is the average atomic number of baryons ($\bar{Z}= 1$ for Hydrogen atom and $\bar{Z}= 1/2$ for general baryonic matter), $m_N$ is nucleon mass  and $M_B$  is the baryon mass. Since the value of number density of $e^+~e^-$ can only be obtained numerically, for simplicity we assume here $n_{e^+e^-} \ll n_B$ and we have numerically checked that this assumption is indeed valid for the values of ${\cal B}$ considered here; namely ${\cal B}=10^{-3}$--$10^{-2}$. In addition, we assume the constant slab approximation with a constant width $\Delta_{lab}$ in the laboratory frame following \citet{1999A&A...350..334R,2000A&A...359..855R}. 

Therefore, the lower bound of transparency radius is 
\begin{equation}
R^{\rm tr} = \left(\frac{\sigma_T}{8 \pi} \frac{M_B }{m_N}\right)^{1/2}.
\label{eq:transp2}
\end{equation}
By substituting Eq.~(\ref{eq:transp2}) in Eq.~(\ref{eq:energy3}), and dividing it by $E_{\rm iso}$, one obtains:
\begin{equation}
\frac{E^{obs}_{\rm P-GRB}}{E_{\rm iso}} = \frac{1}{2} a T^4_{obs} \Gamma^4(1-v/c)^3  \sigma_T \frac{{\cal B}}{m_N c^2} \Delta_{lab}\, ,
\label{eq:energy4}
\end{equation}
where we used the fact that, by definition, ${\cal B}\equiv M_B c^2/E_{\rm iso}$. 

Using the fact that:
\begin{equation}
1-v/c=\frac{1}{(1+v/c)\Gamma^2}\simeq\frac{1}{2\Gamma^2}\, ,
\end{equation}
where we assumed $v/c \sim 1$, that is certainly accurate at the transparency of the \textcolor{black}{PEMB pulses}, we have that:
\begin{equation}
\frac{E^{obs}_{\rm P-GRB}}{E_{\rm iso}} =  \frac{a T^4_{obs}}{16\Gamma^2}\sigma_T \frac{{\cal B}}{m_N c^2} \Delta_{lab}\, .
\label{eq:energy4c}
\end{equation}

From the total energy conservation we have that:
\begin{equation}
E_{\rm iso} = E^{obs}_{\rm P-GRB} + E_{\rm Kinetic}\, ,
\label{eq:energy5}
\end{equation}
therefore
\begin{equation}
1 = \frac{E^{obs}_{ \rm P-GRB}}{E_{\rm iso}} + \frac{E_{\rm Kinetic}}{E_{\rm iso}}\, 
\label{eq:energy6}
\end{equation}
where $E_{\rm Kinetic}$ is the kinetic energy of the baryonic \textcolor{black}{PEMB pulses}:
\begin{equation}
E_{\rm Kinetic} = (\Gamma -1 ) M_B c^2\, .
\label{eq:energy7a}
\end{equation}
By substituting Eq.~(\ref{eq:energy7a}) in Eq.~(\ref{eq:energy6}) we have
\begin{equation}
{\cal B} = \frac{1}{\Gamma -1}\left(1-\frac{E^{obs}_{\rm P-GRB}}{E_{\rm iso}}\right),
\label{eq:energy7}
\end{equation}
or, equivalently:
\begin{equation}
\Gamma = 1+ {\cal B}^{-1}\left(1-\frac{E^{obs}_{\rm P-GRB}}{E_{\rm iso}}\right).
\label{eq:energy8}
\end{equation}
The radius of transparency, $R^{\rm tr}$, is given by Eq.~(\ref{eq:transp2}) in this theoretical approach:
\begin{equation}
    R^{\rm tr} =\left(\frac{\sigma_T}{8\pi}\frac{{\cal B} E_{\rm iso}}{m_Nc^2}\right)^{1/2}.
\label{eq:arri}
\end{equation}

In general, from Eqs.~(\ref{eq:energy4c}) and (\ref{eq:energy7}), the values of ${\cal B}$ and $\Gamma$ can be estimated by the  values of $E^{obs}_{\rm P-GRB}/E_{\rm iso}$, $T_{obs}$ and $\Delta_{\rm lab}$. Also, having $E_{\rm iso}$ and ${\cal B}$, we can obtain the transparency radius from Eq.~(\ref{eq:arri}).

\begin{figure*}
\centering
[a]\includegraphics[width=0.41\hsize,clip]{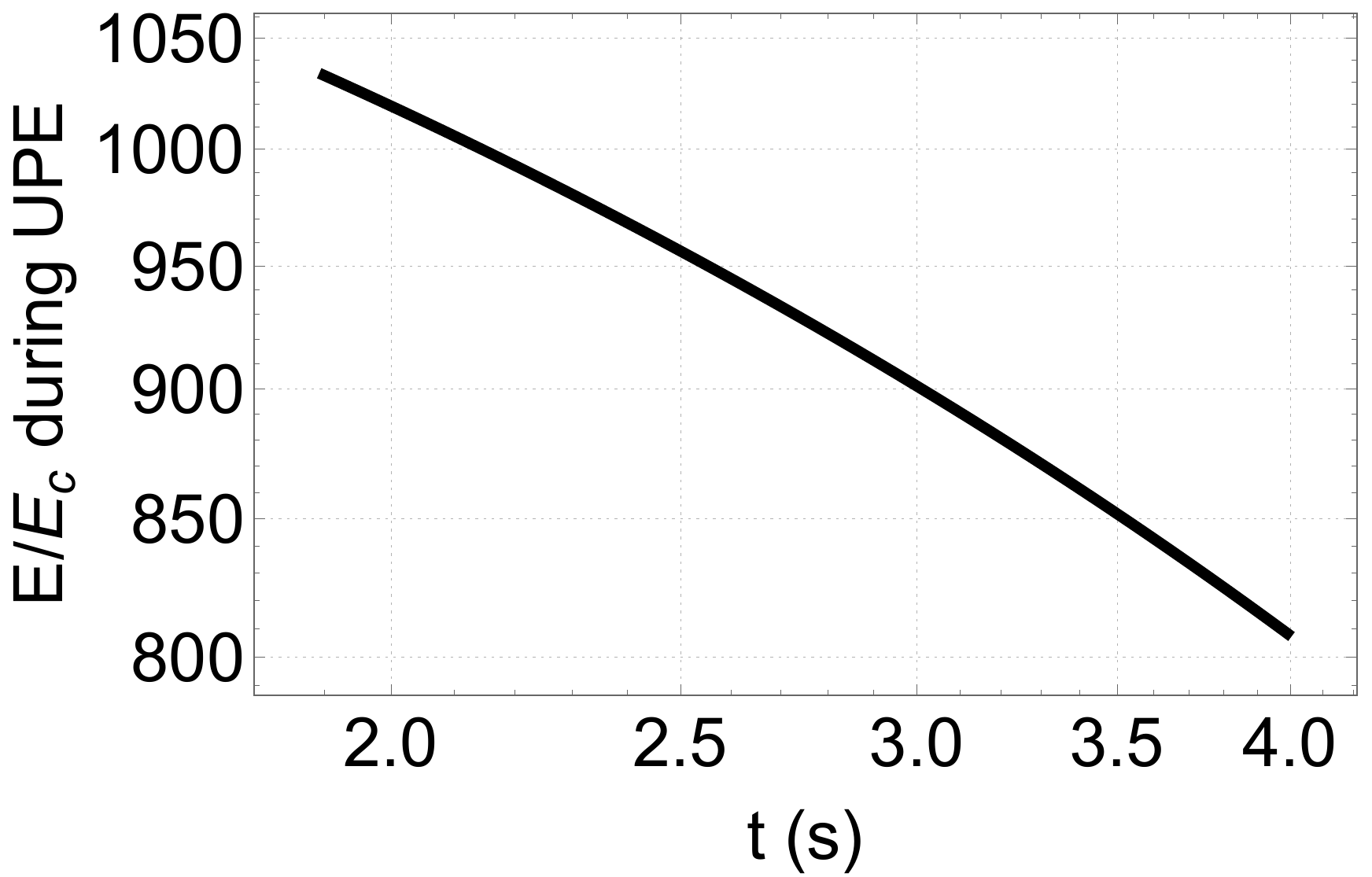}
[b]\includegraphics[width=0.4\hsize,clip]{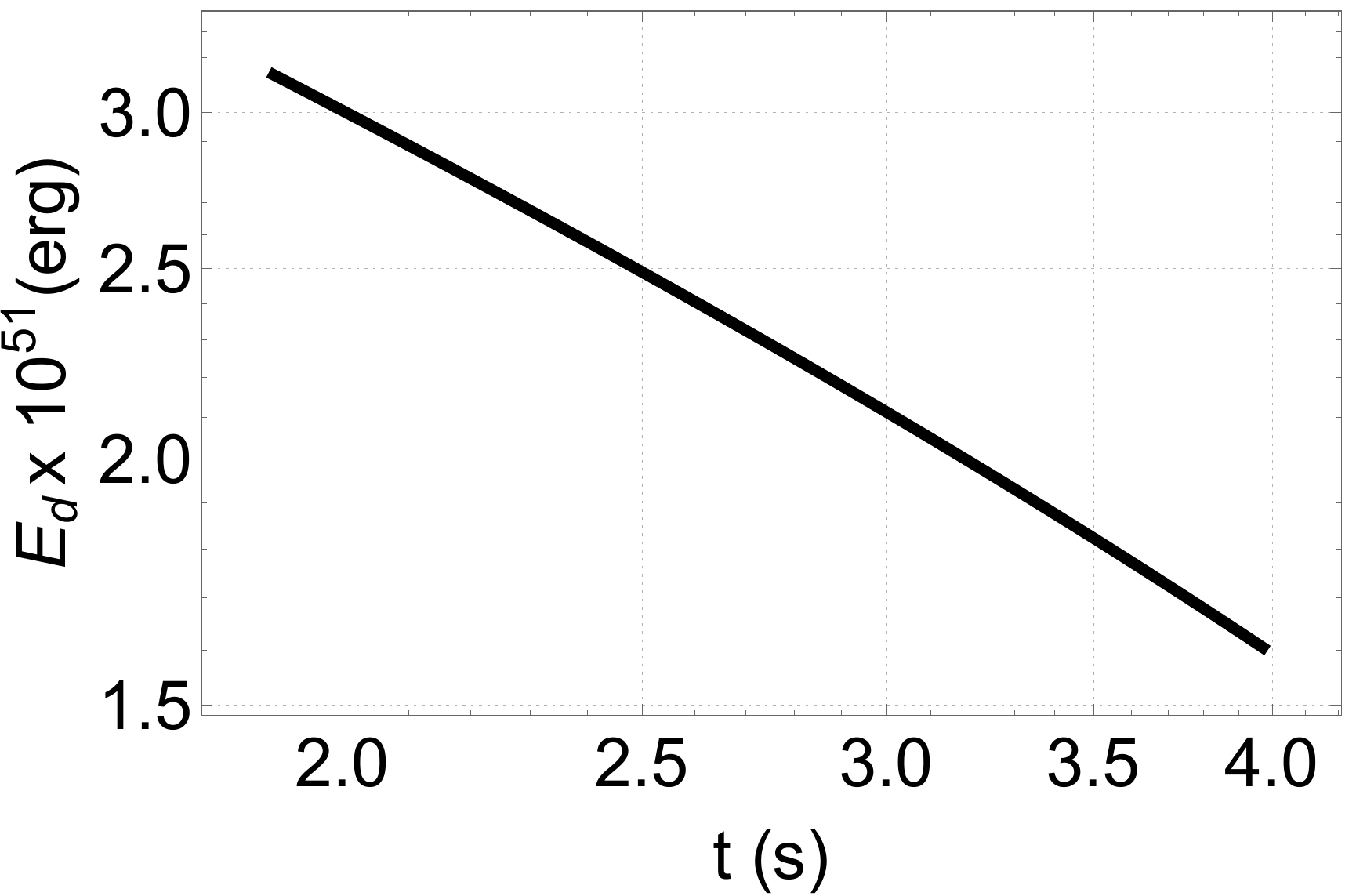}
[c]\includegraphics[width=0.4\hsize,clip]{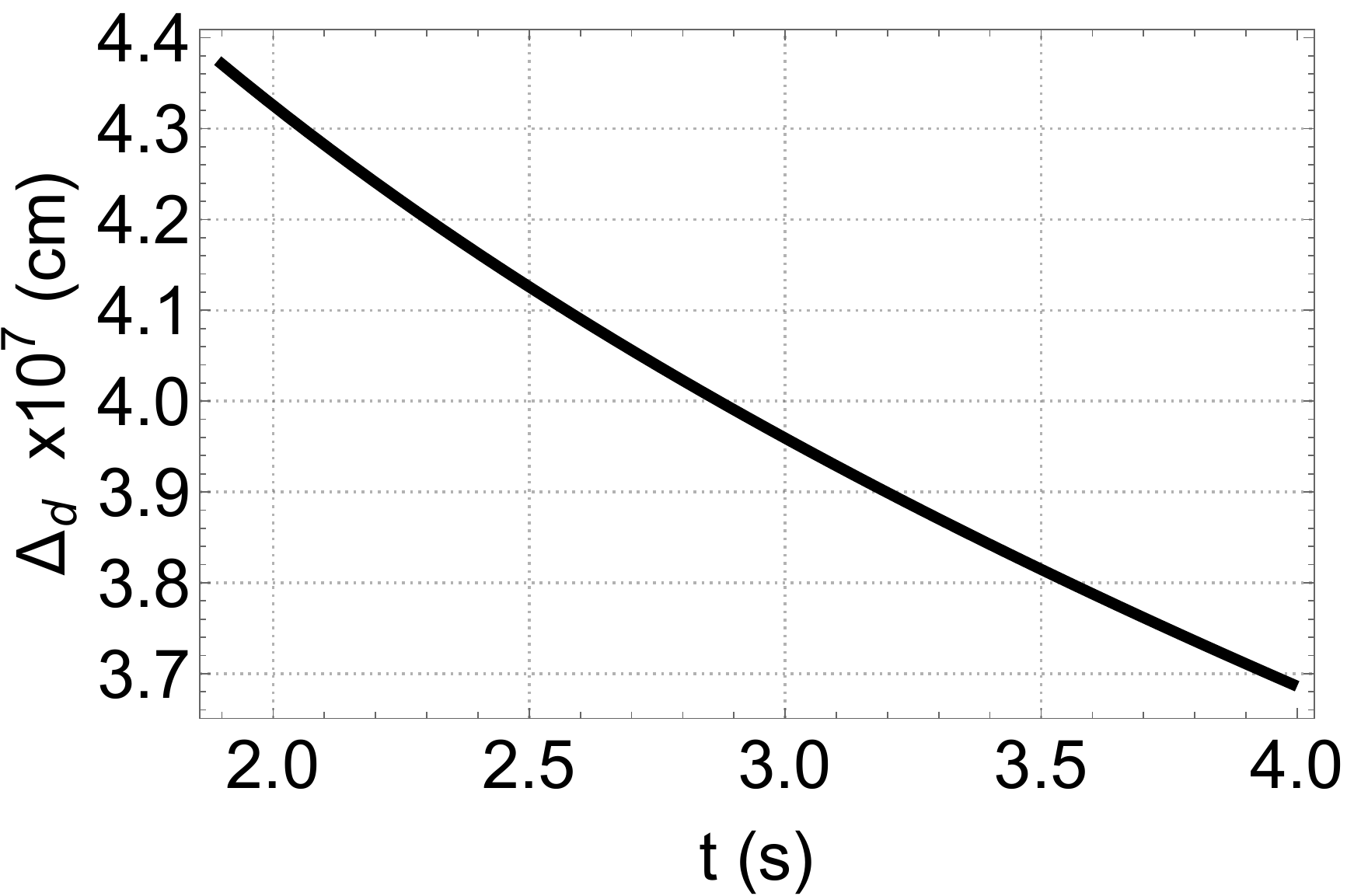}
[d]\includegraphics[width=0.4\hsize,clip]{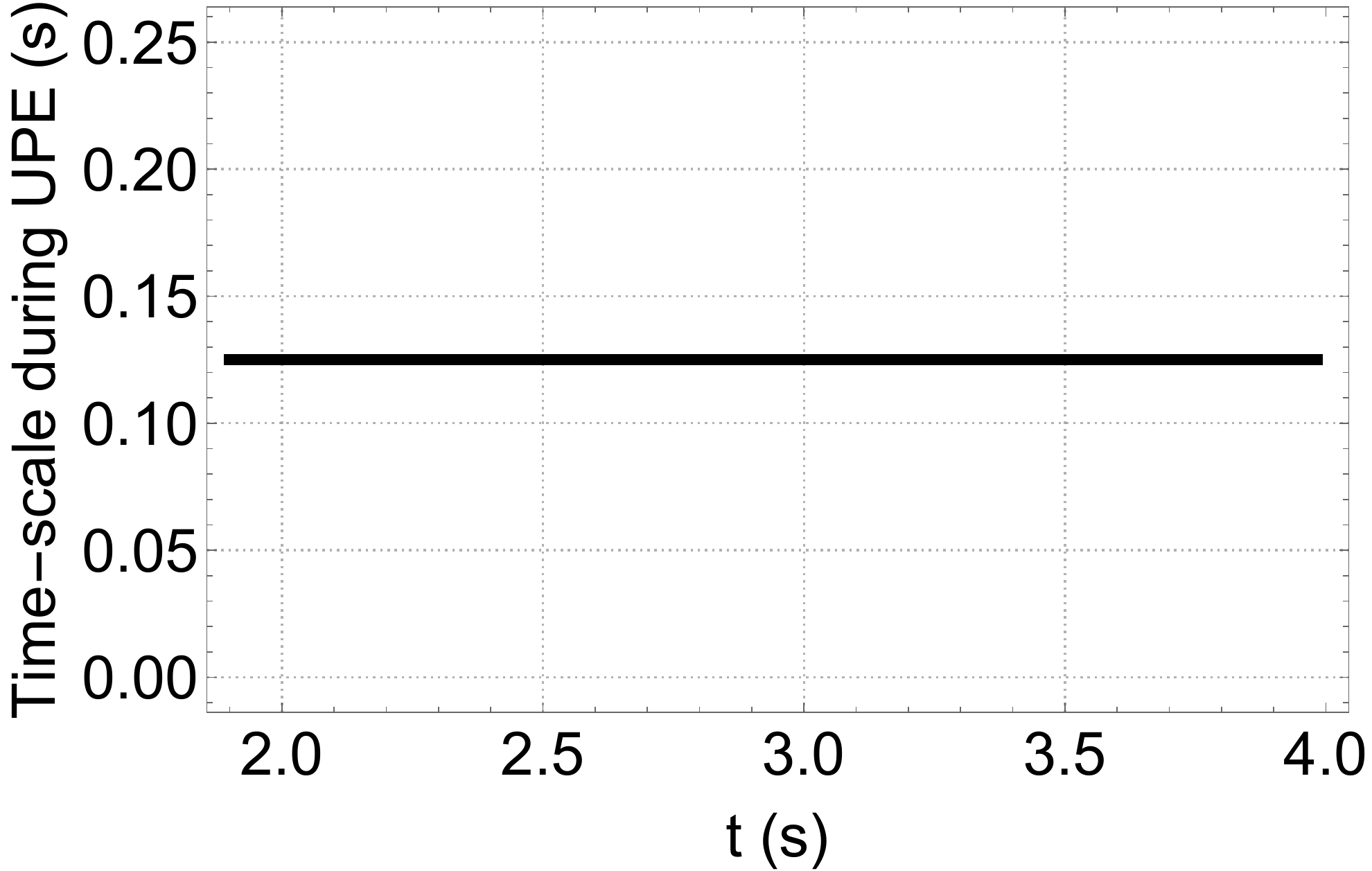}
[e]\includegraphics[width=0.41\hsize,clip]{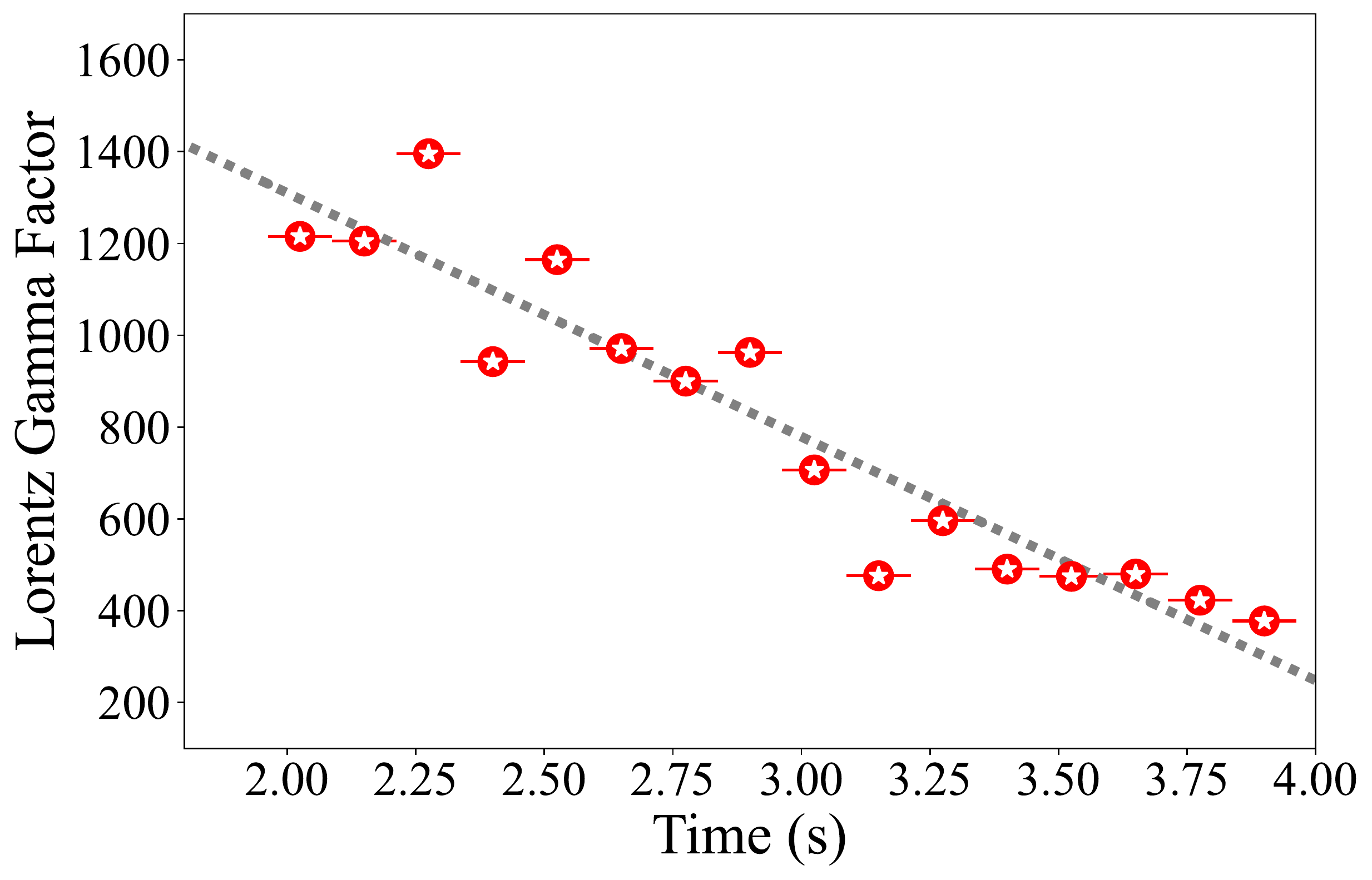}
[f]\includegraphics[width=0.41\hsize,clip]{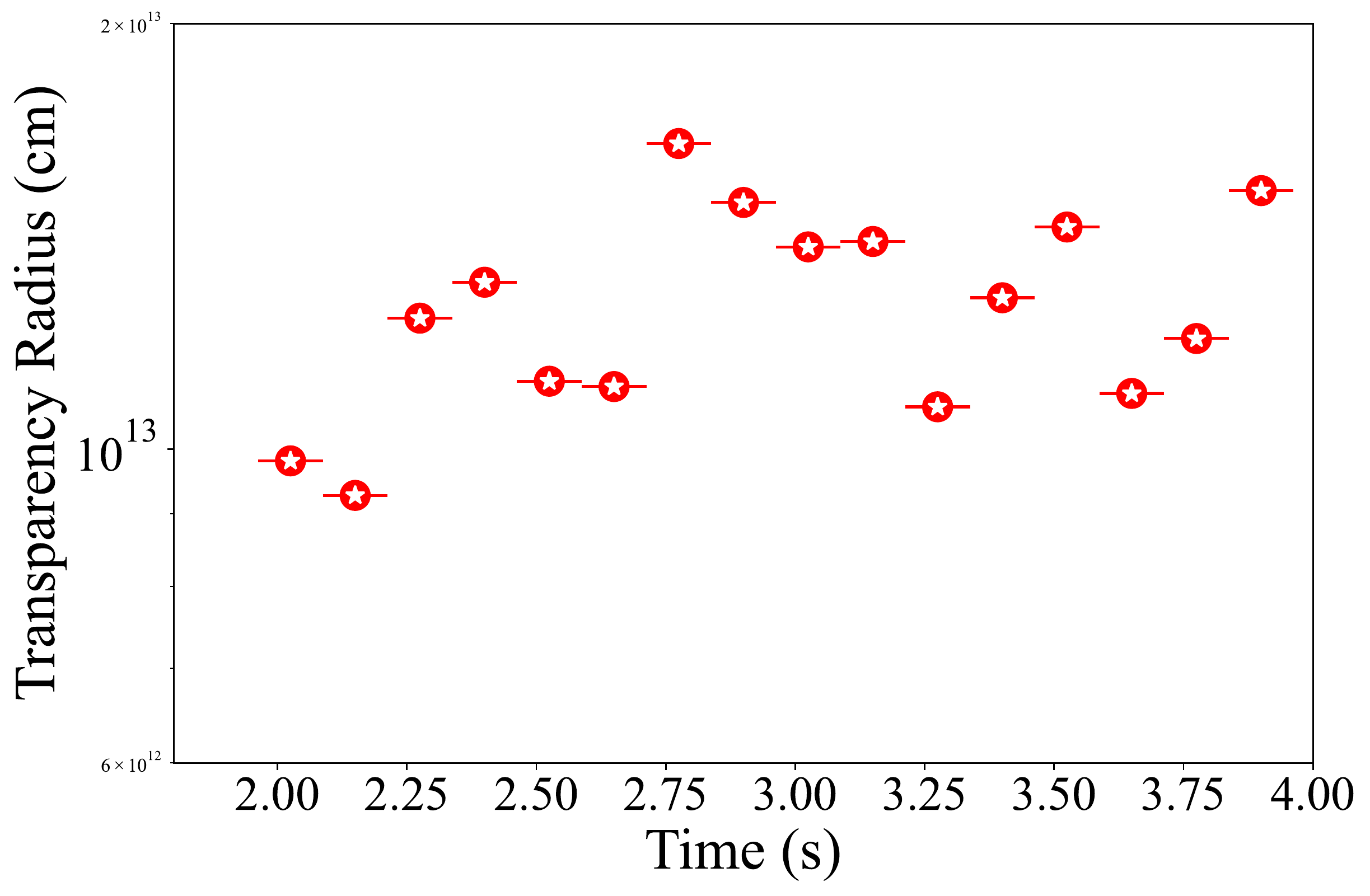}
\caption{\label{fig:Lrentz1051} The parameters of \emph{inner engine}  and transparency condition as  a function of rest-frame time for GRB 190114C during the UPE phase, namely in the rest-frame time interval $t_{\rm rf}=1.9$--$3.99$~s, obtained from a time-resolved analysis down to a $\Delta t=0.125$~s time resolution reported in Table~\ref{tab:table}. \textbf{(a):} The electric field  during UPE phase which is clearly overcritical. \textbf{(b):} The energy of dyadoregion  during the UPE phase obtained from Eq. \ref{Eemxi}. \textbf{(c):} The width of dyadoregion obtained from Eq. \ref{eq:width}. \textbf{[d]:} Timescale of radiation during the UPE phase. \textbf{(e):} The decrease of the Lorentz gamma factor, $\Gamma$, as a function of rest-frame time. \textbf{(f):} The evolution of transparency radius in the UPE phase of GRB 190114C. All values are obtained \textcolor{black}{for magnetic field of}, $B_0=1.8 \times 10^{17}$~G, calculated in Sec. \ref{sec:mag-fifth}.}
\end{figure*}
%
\section{The magnetic field inferred from the $\Delta \MakeLowercase{t}=0.125$~s time resolved spectral analysis}\label{sec:mag-fifth}

As a \textcolor{black}{specific} example, we calculate the magnetic field and \textcolor{black}{transparency parameters} for the $\Delta t_{\rm rf}= 0.125$~s time resolved interval, \textcolor{black}{namely the fifth iteration} analysis reported in Sec.~\ref{sec:heir}. \textcolor{black}{Therefore,} the UPE phase \textcolor{black}{is assumed to be composed of} $16$ \textcolor{black}{expanding PEMB pulses} \textcolor{black}{emitting an} average \textcolor{black}{isotropic energy of} $E_{\rm iso}\sim  10^{52}$~erg, \textcolor{black}{with radiation time scale of} $\tau_q=0.125~$s, as reported in Table~\ref{tab:table}. 

\textcolor{black}{Therefore, from 1) the ratio $E_{\rm \rm P-GRB}/E_{\rm iso}= 0.3$, and \textcolor{black}{2)} $E_{\rm \rm P-GRB}=E_{(r_+,r_{\rm d})}$, the electromagnetic energy stored in each expanding \textcolor{black}{PEMB pulse} should be $E_{(r_+,r_{\rm d})}=0.3\times E_{\rm iso}$. Consequently, from Eq.~(\ref{Eemxi}) and the value of mass and spin parameter of the BH, the magnetic field needed to fulfill this energetic is $B_0= 1.85 \times 10^{17}$~G.} The Lorentz factor, \textcolor{black}{$\Gamma \sim 1000$}, \textcolor{black}{the baryon load}, ${\cal B} \sim 2 \times 10^{-3}$, and \textcolor{black}{the radius of transparency}, $ R^{\rm tr} \sim 10^{13}~\rm cm$, are obtained using: 
\begin{itemize}
    \item the isotropic energy of each time interval, $E_{\rm iso} \sim 10^{52}$~erg;
    \item the ratio of black body energy to isotropic energy $E^{\rm obs}_{\rm P-GRB}/E_{\rm iso}\sim 0.3$;
    \item
    the value of black body temperature in keV reported in Table~\ref{tab:table};
    \item the width of the dyadoregion at decoupling, $\Delta_{\rm lab}=\Delta_{\rm d}$ obtained from Eq.~(\ref{eq:width}) for magnetic field of $B_0=1.85 \times10^{17}~$G.
\end{itemize}

\textcolor{black}{The results are} shown in Fig.~\ref{fig:Lrentz1051}. \textcolor{black}{It is appropriate to notice that the magnetic field of $B_{0}=1.85 \times 10^{17}$~G, obtained from the $\Delta t_{\rm rf}= 0.125$~s time-resolved analysis, does not fulfill the boundary condition of the UPE phase,  $|{\bf E}|=E_c$ at $t_{\rm rf}=3.99$ s. In the next section, we calculate the lowest limit of magnetic field and the minimum repetition time which fulfill the required boundary condition $|{\bf E}|=E_c$ at $t_{\rm rf}=3.99$ s.}

\section{The lower limit of magnetic field during the UPE phase}\label{sec:mag-low}

 Having determined the boundary value of the magnetic field at  $t_{\rm rf}=3.99$~s to be $B_0= 3.9 \times 10^{10}$~G, we must now require that  at $t_{\rm rf}>3.99$~s the electric to be \textcolor{black}{under}critical, and overcritical inside the UPE phase. In section~\ref{sec:massupe},  we have determined the overall behavior of the mass and spin of BH during the UPE since the moment of the formation of BH; see Fig.~\ref{massspinupe}.

\begin{figure*}
\centering
[a]\includegraphics[width=0.45\hsize,clip]{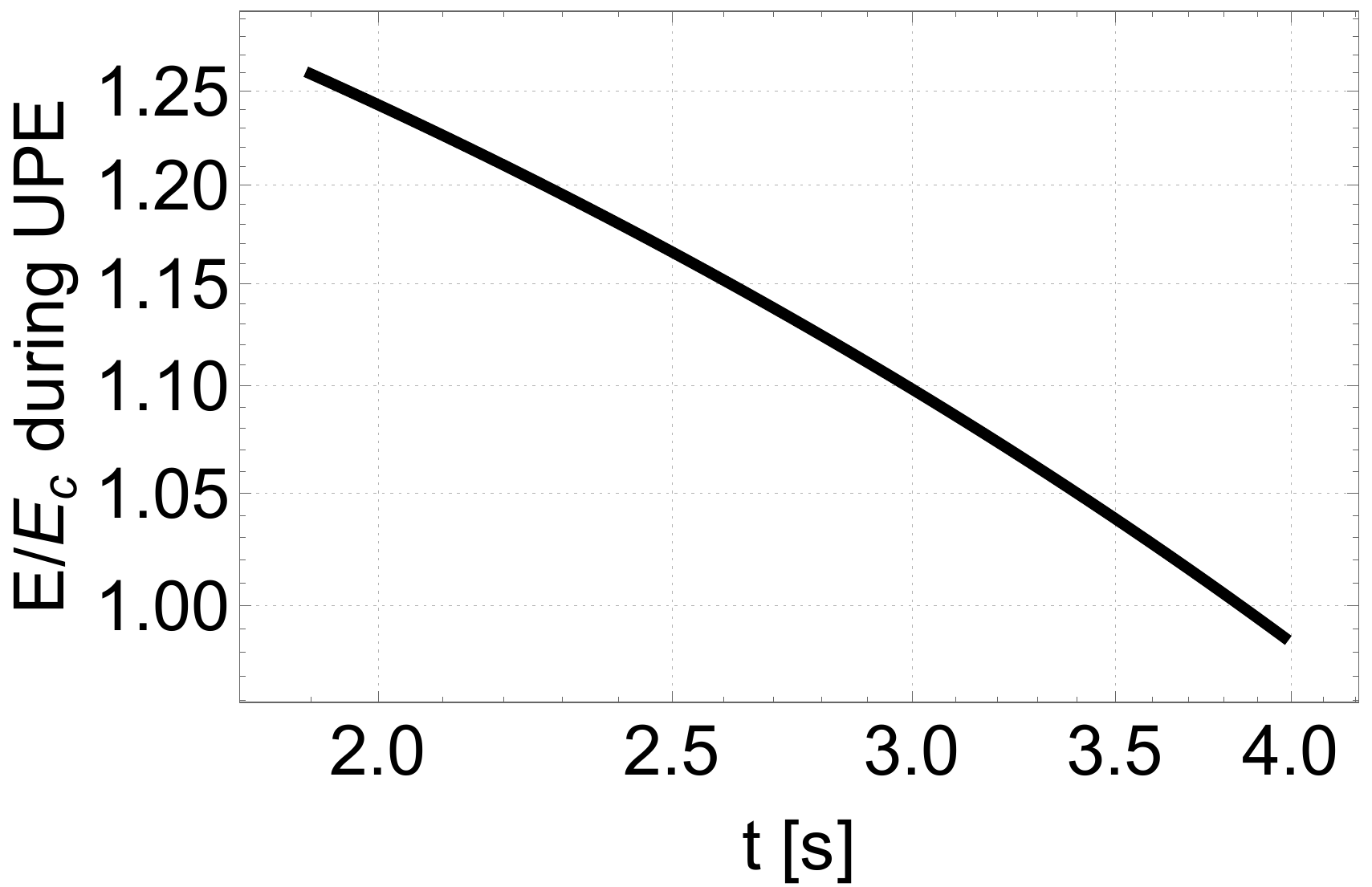}
[b]\includegraphics[width=0.45\hsize,clip]{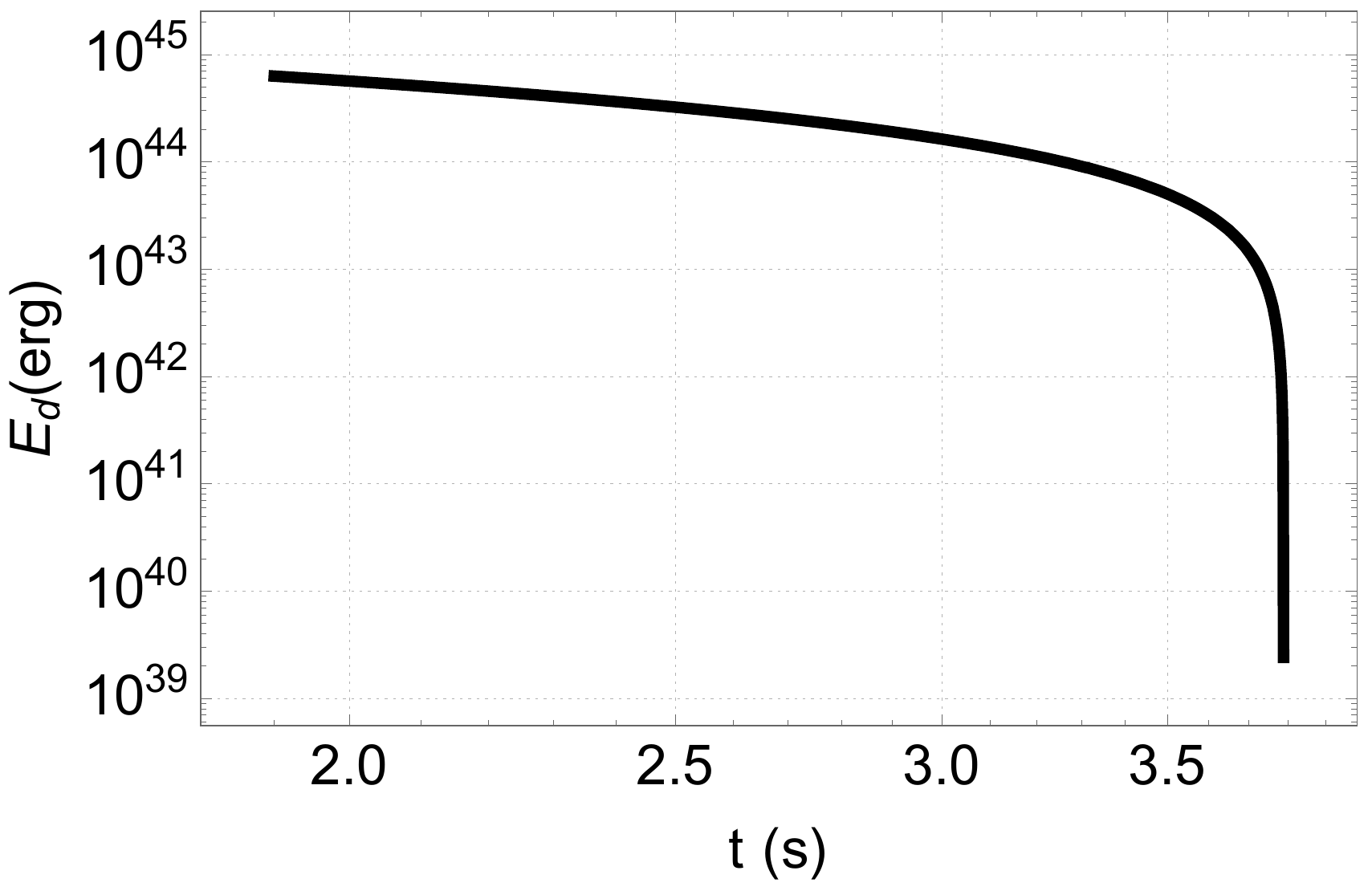}
[c]\includegraphics[width=0.45\hsize,clip]{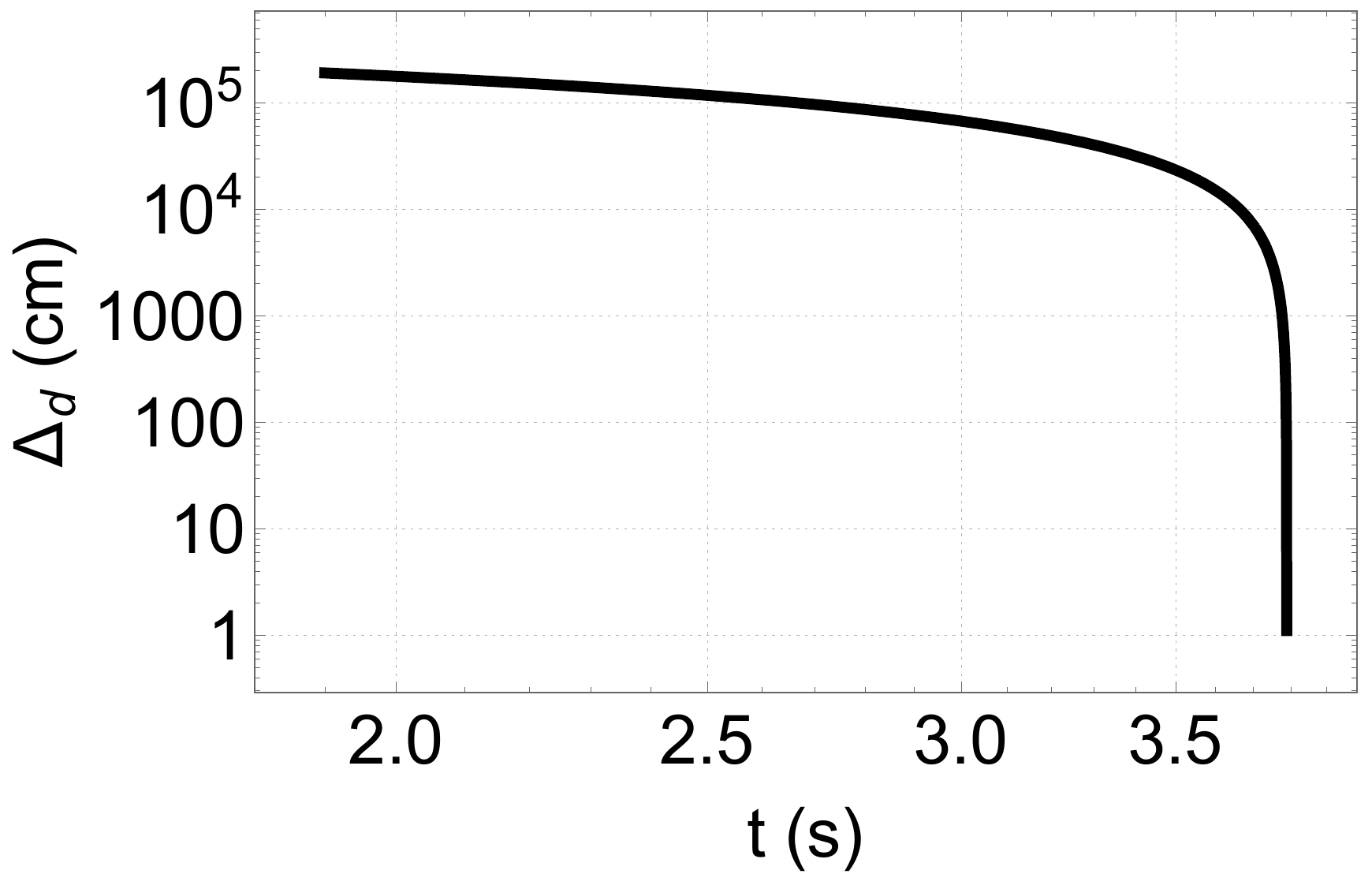}
[d]\includegraphics[width=0.45\hsize,clip]{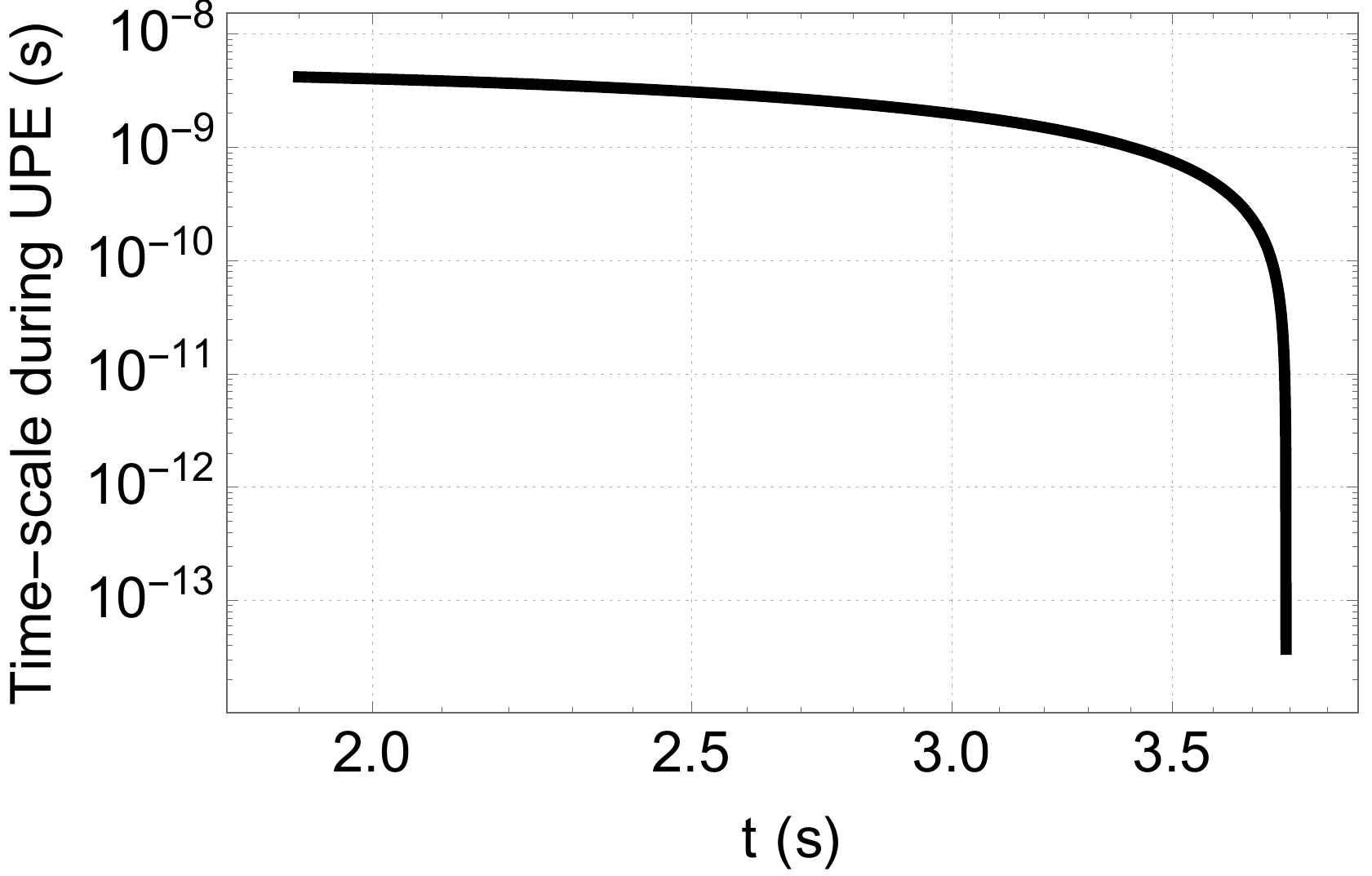}
[e]\includegraphics[width=7.1cm]{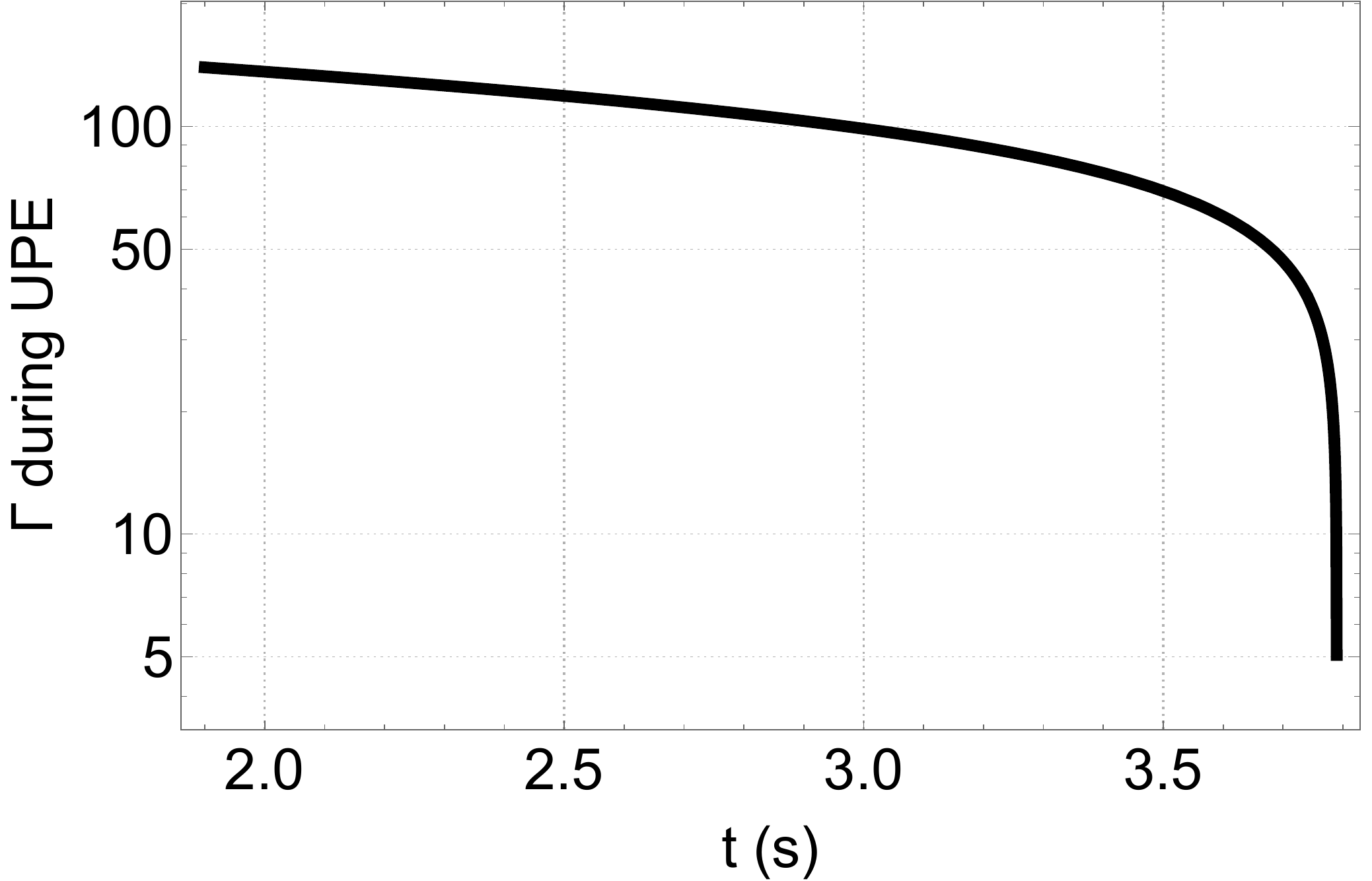}
[f]\includegraphics[width=8.5cm]{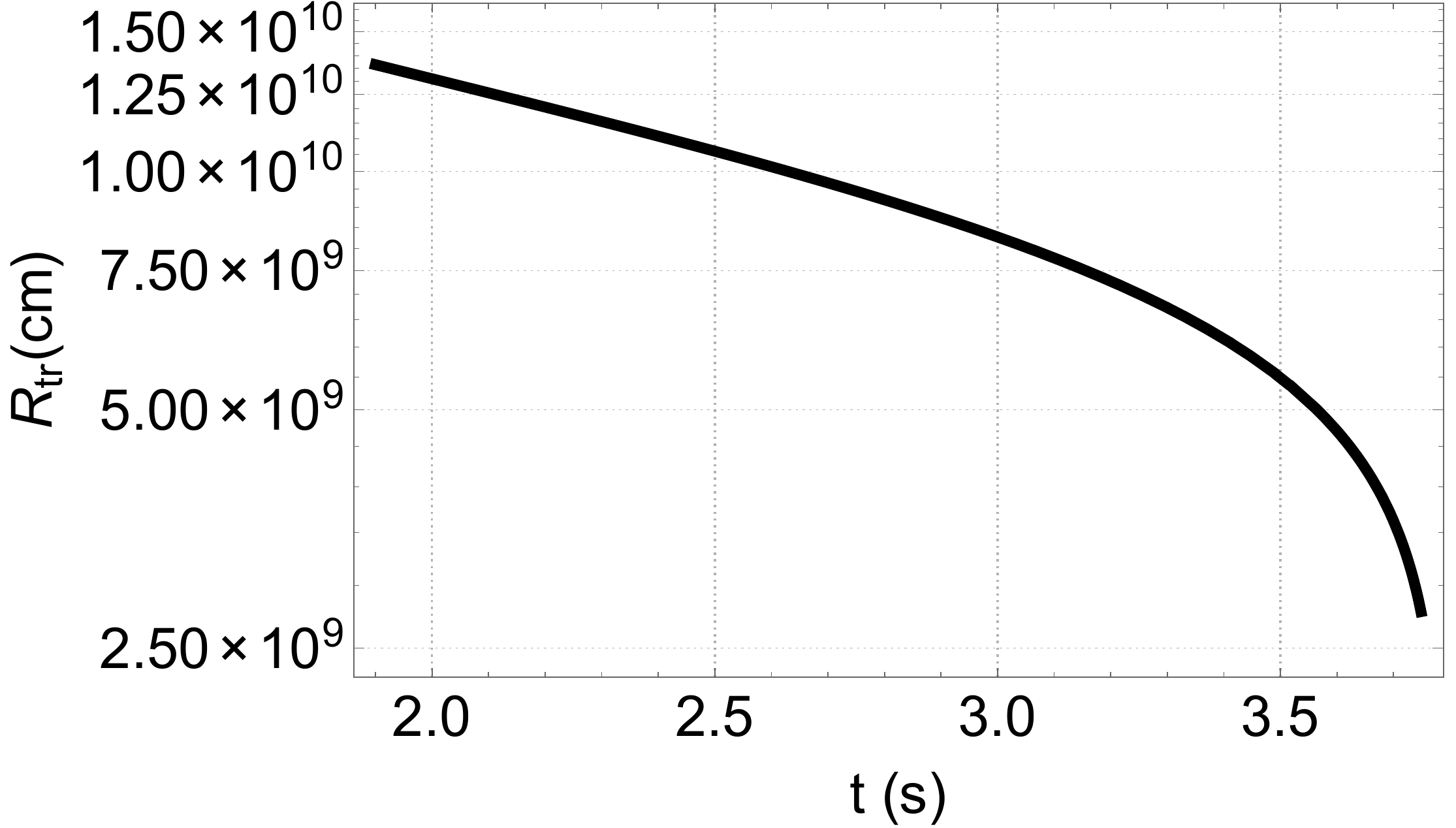}
\caption{
\textcolor{black}{The parameters of \emph{inner engine}  and transparency point, obtained for $B_0=2.3 \times 10^{14}$~G, as a function of rest-frame time for GRB 190114C during the UPE phase, namely in the rest-frame time interval $t_{\rm rf}=1.9$--$3.99$~s.} \textbf{(a):} The electric field during UPE phase \textcolor{black}{which} is clearly overcritical and reaches its critical value at the end of UPE phase \textcolor{black}{(t$_{\rm rf}=$3.99~s)}. \textbf{(b):} The energy of dyadoregion  during the UPE phase obtained from Eq. (\ref{Eemxi}). \textbf{(c):} The width of dyadoregion obtained from Eq. (\ref{eq:width}) which tends to zero at the end of UPE phase, indicating that the number of $e^+ e^-$ pairs are suppressed and the UPE phase is \textcolor{black}{over}. \textbf{(d):} Repetition time scale of the \emph{inner engine}  during the UPE phase obtained from Eq. (\ref{tauinner}). \textbf{(e):} The decrease of the Lorentz factor, $\Gamma$, as a function of rest-frame time. This indicates the fact that $\Gamma$ tends to unity for the last layers which confirms the the end of UPE is reached. \textbf{(f):} The evolution of transparency radius; see Sec. \ref{sec:mag-low}. }
\label{fig:eduringupe}
\end{figure*}

We set the value of $B_0$ in the UPE phase, i.e. at times $t_{\rm rf}< 3.99$~s, such that the electric field therein is overcritical. The lower limit of the magnitude of magnetic field  is determined in a way that in Eq.~(\ref{eq:ER2}), $|E_{r_+}|= E_c$  at the end of the UPE phase; at $t_{\rm rf}= 3.99$~s. For BH mass and spin parameter at the end of UPE, it implies a magnetic field of $\beta=B_0/B_c = 5.1$ or $B_0=2.3 \times 10^{14}$~G; see Fig.~\ref{fig:eduringupe}.

For $B_0=2.3 \times 10^{14}$~G and at the moment which BH is formed, namely $t_{\rm rf} = 1.9$~s, $\lambda\approx 4.7 \times 10^{-5}$ (from Eq.~\ref{eq:lamda}), which leads to $r_{d}=1.15~r_+$. Having these values, the energy of dyadoregion at $t_{\rm rf} = 1.9$~s is $E_{d} \approx 6.27 \times 10^{44}$~erg. The evolution of energy of dyadoregion is shown in Fig.~\ref{fig:eduringupe} [b].

The evolution of characteristic width of the dyadoregion is shown in Fig.~\ref{fig:eduringupe} [c]. At $t_{\rm rf} = 3.99$~s, the extent of the dyadoregion tends to zero confirming that no enough $e^+~e^-$ pairs are created and the UPE phase is finished.

\subsection{The transparency condition obtained from the lower limit of magnetic field, $B_0=2.3 \times 10^{14}$~G}\label{sec:trans}

For $B_0=2.3 \times 10^{14}$~G, each expanding \textcolor{black}{PEMB pulse}; which \textcolor{black}{are produced via vacuum polarization and self-expanded with} different Lorentz factors, has an isotropic energy $\sim 10^{45}$~erg obtained from Eq.~(\ref{Eemxi}); see Fig.~\ref{fig:eduringupe} 
The total isotropic energy of \textcolor{black}{the UPE phase is} $E^{\rm UPE}_{\rm iso}=1.47 \times 10^{53}$~erg, therefore, this phase consists of $\sim 10^{8}$ impulses in the time interval $1.9$--$3.99$~s. Radiation from each one of these \textcolor{black}{PEMB pulse} can be interpreted as a blackholic quantum introduced in \citet{2020EPJC...80..300R}.

As expressed in Sec.~\ref{sec:12}, the key parameters for calculating the transparency radius of each impulse are: (1) its isotropic energy, $E_{\rm iso}$, (2) the blackbody to isotropic energy ratio, $E^{\rm obs}_{\rm P-GRB}/E_{\rm iso}$, (3) the blackbody temperature, ($T_{\rm obs}$), and finally (4) the width $\Delta_{\rm lab}=\Delta_{\rm d}$. 

From the \emph{inner engine}  theory, as presented in the previous subsection, for each impulse we have $E_{\rm iso}\sim 10^{45}$~erg and the width of the dyadoregion at decoupling is $\Delta_{\rm d}= 1.9 \times  10^{5}$~cm. From the hierarchical structure of UPE phase in this GRB presented by Eq.~\ref{eq:ratio}, we have $E^{\rm obs}_{\rm P-GRB}/E_{\rm iso}\sim 0.3$ and the temperature $k T_{\rm obs}\sim 150$~keV.

With these values of $E^{\rm obs}_{\rm P-GRB}/E_{\rm iso}$, $\Delta_{\rm d}$, $T_{\rm obs}$, and $E_{\rm iso}$, we obtain via Eqs.~(\ref{eq:energy4c}), (\ref{eq:energy7}) and (\ref{eq:arri}), the transparency radius of
\begin{equation}
 R^{\rm tr} = 9.4\times 10^9~\rm cm,
\end{equation}
the baryon load parameter
\begin{equation} 
{\cal B} = 5.1 \times 10^{-3},
\end{equation}
and finally the Lorentz factor of
\begin{equation} 
 \Gamma \approx 139.
\end{equation}

We have checked that these estimated values are in good agreement with the corresponding ones obtained from the numerical simulation of the \textcolor{black}{PEMB pulses} evolution. The corresponding values from the numerical simulation are: $R^{\rm tr}= 9.3\times 10^9$~cm, temperature $k T=150$~keV, and Lorentz factor $\Gamma \sim 140$. 

The evolution of the Lorentz gamma factor, $\Gamma$ is shown in Fig.~\ref{fig:eduringupe}(e) indicating the fact that $\Gamma$ tends to $\sim $ 4 for the last shell which confirms the end of UPE is reached.

The evolution of transparency radius in the UPE of GRB 190114C, using the exact numerical values of energy and width of dyadoregion are also shown in Fig.~\ref{fig:eduringupe} (f).

\subsection{The repetition time of sequence of the blackholic quanta  }\label{sec:6}

We now study the timescale of each blackholic quanta in which the system starts over. The new value of the electric field is set by the new values of the BH angular momentum and mass, $J = J_0-\Delta J$ and $M = M_0-\Delta M$, keeping the magnetic field value constant $B_0$ in which at $t_{\rm rf}=1.9$ s, i.e.
\begin{equation}
\frac{\Delta J}{J} \approx \frac{\Delta M}{M}= 3.1\times 10^{-9}.
\end{equation}

Regarding the presence of the baryon load obtained from Eq.~(\ref{eq:energy7}) in the acceleration process, we infer from the MeV luminosity, the evolution of the timescale $\tau_q(t)$  of the blackholic quantum by requiring it to explain the MeV emission, i.e.:
\begin{equation}
\label{eq:tauMeV}
 L_{\rm MeV}=\frac{[1-{\cal B}(\Gamma-1)]~E_{(r_+(t),r_d(t))}}{\tau_q (t)}.
\end{equation}

In fact, the effect of baryon load is $(1-{\cal B}~\Gamma)\approx 0.3$. Therefore, we obtain for the timescale
\begin{equation}\label{tauinner}
  \tau_q(t)= \frac{0.3~E_{(r_+(t),r_d(t))}}{ L_{\rm MeV}}.
\end{equation}
where the $E_{(r_+(t),r_d(t))}$ is the energy of dyadoregion obtained from Eq.~(\ref{Eemxi}), determined from the new values of $J$ and $M$ for each blackholic quanta and $L_{\rm MeV}$ is the MeV luminosity obtained from best fit represented by Eq.~(\ref{Mevluminosity}); the evolution of the blackholic timescale is shown in Fig.~\ref{fig:eduringupe} (d).

\textcolor{black}{\subsection{The approach of the $|{\bf E}|=E_c$ at the UPE phase boundary }\label{sec:boundary}}

From the above theoretical derivation, we can explicitly see that, for an iteration, such that the duration of each elementary process of the nth iteration is $10^{-9}$~s, namely after $10^{9}$ iterations, the physical model can be consistently implemented, deriving the necessary parameters characterizing the process, namely the energy of each \textcolor{black}{PEMB pulse}, the baryon load, the Lorentz factor, and the radius at transparency. \textcolor{black}{From Fig.~\ref{fig:eduringupe}, it becomes clear that after $t_{\rm rf}\sim$ 3.7 s, the emission of quanta by the QED process becomes not effective and the classical regime is soon approached.}

\textcolor{black}{The lowest limit of the magnetic field to reach $|{\bf E}|=E_c$ occurs in an \emph{inner engine}  composed of a Kerr BH of initial mass of $M=4.53 M_{\odot}$ and $\alpha=0.51$, immersed with a uniform magnetic of $B_0=2.3 \times 10^{14}$~G with a radiation timescale of $\sim 10^{-9}$~s.}

\textcolor{black}{Indeed, the decrease of the magnetic field from $\beta=5.1$ to $\beta=8.9\times 10^{-4}$ at $t_{\rm rf}= 3.99$~s, can be explained as the result of the induced current created by pairs in the inward electric field, which screens the original magnetic field. This is a very interesting process that has consequences in different astrophysical scenarios. Therefore, we here limit ourselves to the above explanation and refer the reader for further details in the dedicated, separated publication \citep{CAMPION2021136562}.}

\textcolor{black}{All the above results: (1) are in perfect agreement with observational data; see Fig.~\ref{fig:lumupe} and, (2) overcome the compactness problem of the UPE phase. It is appropriate to mention all these results have been obtained guided by the hierarchical structure of the UPE phase. }

\textcolor{black}{\section{Comparison with other approaches}\label{sec:comparison}}

\textcolor{black}{The magnetohydrodynamics of plasma accretion onto the Kerr BH was first addressed in \citet{1975PhRvD..12.2959R} assuming he infinite conductivity condition, $F_{\rm \mu \nu}U^{\rm \nu}=0$ implying $\mathbf{E}\cdot \mathbf{B} =0$. In view of zero net charge on the surface of the BH, no process of energy extraction, neither by vacuum polarization nor by electromagnetic process was there possible \citep{Cherubini:2018fba}.}

\textcolor{black}{\citet{1977MNRAS.179..433B} returned on the same process and in order to overcome the difficulty of extracting energy and they introduced, in analogy with pulsar, the presence ``\textit{gaps}'' \citep{1971ApJ...164..529S, 1975ApJ...196...51R}.} 

\textcolor{black}{\citet{1982MNRAS.198..339T}, following \citet{1973PhRvD...8.3259H}, calculated the surface charge induced on the horizon of the Kerr BH immersed in the magnetic field in the Papapetrou-Wald solution \cite{1974PhRvD..10.1680W}. \citet{2000NCimB.115..751M} explicitly manifested that the Papapetrou-Wald solution \cite{1974PhRvD..10.1680W} implies $\mathbf{E} \cdot \mathbf{B} \neq 0$, and identified that the induced surface charge implies a quadrupolar distribution of electric field around the BH. These results were confirmed in \citep{2008ASSL..355.....P}. Applying  these works to the case of GRBs, it has been shown that the mathematical Papapetrou-Wald solution can be used in order to describe the \textit{inner engine} of a GRB 130427A \citep{2019ApJ...886...82R}, which presents mechanism to extract the rotational energy of the Kerr BH. The process which occurs in the  undercritical field regime leads to the emission of synchrotron radiation in the GeV domain as well as ultrahigh energy cosmic rays (UHECRs) \citep{2021A&A...649A..75M}. The synchrotron emission of the \textit{inner engine}  occurs near the BH horizon and is emitting in blackholic quanta \citep{2020EPJC...80..300R}.}

\textcolor{black}{The extrapolation to overcritical field regime, presented in this paper, leads to the explanation of the MeV radiation during the UPE phase.} 

\textcolor{black}{In recent years, in parallel to the theoretical progresses in the field, computer simulations were also developed. These simulations point the fact that present plasma in any energy extracting scheme would screen the background electric field of the vacuum solution of Papapetrou-Wald from the magnetosphere; see e.g.  \citet{2005MNRAS.359..801K,2019PhRvL.122c5101P}. These simulation mainly address the physics of active galactic nuclei (AGNs) and particularly attentive has been the review of their theoretical models indicated in \citet{2005MNRAS.359..801K}. Their choice of parameters and physical processes are quite different from the ones we have used for the GRB analysis. In our GRB approach we have been guided by the theoretical explanation of a vast number of observations obtained from: 1) the unprecedented time-resolved spectral analysis of the UPE phase; 2) the power-law MeV luminosity observed by Fermi-GBM; and 3) the power-law GeV luminosity observed by Fermi-LAT. This allows us to identify the physical processes and parameters which had to be fulfilled in order to obtain the detailed acquired data. Their choice of parameters enforce $\mathbf{E} \cdot \mathbf{B} = 0$ condition so different from $\mathbf{E} \cdot \mathbf{B} \neq 0$ which has allowed us to obtain our results.}


\textcolor{black}{In our model, the magnetic field is left over by the collapse of the accreting NS to the BH, rooted in the surrounding material, and the electric field is created by the gravitomagnetic interaction of the spacetime rotation with the present magnetic field; see, e.g., \citet{2020ApJ...893..148R}. Following this procedure, and since the electric field is assumed to be overcritical, in a very short timescale $\sim \hbar/(m_e c^2) \approx 10^{-21}$ s, much shorter than any electromagnetic process, a dyadoregion originate dominated by the  high density and high pressure of the neutral $e^+e^-\gamma$ plasma \cite{2010PhR...487....1R}.}


\textcolor{black}{The optically thick pair electromagnetic-baryon (PEMB) pulse self-accelerates to the ultrarelativistic regime and finally reaches the transparency point at the radius of $\sim 10^{10}$~cm. These classical results were obtained thanks to a collaboration with Wilson at Lawrence Livermore National Laboratory \citep{1999A&A...350..334R, 2000A&A...359..855R}}


\textcolor{black}{As soon as the BH is formed, the first and the most efficient process in action to produce the $e^+e^-$ plasma and, consequently decreasing the rotational energy of BH, occurs through the Schwinger critical field pair production. Since an overwhelming amount of pair plasma is created in quantum timescales, the plasma expansion by its internal pressure starts well before any electric field screening. }

\textcolor{black}{This process takes a fraction of angular momentum of the Kerr BH. The BH then is left with a slightly smaller angular momentum $J^* = J- \Delta J$, with $\Delta J/J\sim 10^{-9}$, being $\Delta J$ the angular momentum, and the same magnetic field which leads to a new electric field created by the space-time rotation. As a result, the system starts a new process in presence of the same magnetic field $B_0$, kept rigorously constant and a new effective charge of $Q^*_{\rm eff}= Q_{\rm eff}-\Delta Q_{\rm eff}$ which $\Delta Q_{\rm eff} =2 B_0 \Delta J$.}

\textcolor{black}{This process continues till the moment that electric field is not overcritical anymore, and after that the sole electromagnetic process is at work.} 
\textcolor{black}{The expanding $e^+e^-\gamma$ plasma sweeps away the matter in the cavity whose density after this process becomes $\sim 10^{-14}$ g cm$^{-3}$, and an undercritical electromagnetic field is left; see \citet{2019ApJ...883..191R}. This low-density ionized plasma is needed to fulfill an acceleration of charged particles leading to the electrodynamical process around a newborn BH. In fact, this density is much below the Goldreich-Julian density $\rho_{\rm GJ}= 8 \times 10^{-12}$ g cm$^{-3}$ obtained for the $B_0 = 3.9 \times 10^{10}~$~G and $M=4.45 M_{\odot}$ and $a=0.41 M$. Moreover, the matter energy density inside the cavity is negligible comparing to the electromagnetic energy density, namely $\rho_M/(|B|^2-|E|^2) \sim 10^{-14}$, while in \citet{2005MNRAS.359..801K} this ratio is $0.05$ or higher.}

\textcolor{black}{It is interesting that the \textit{inner engine} operates as well in the supermassive BHs in active galactic nuclei in the $|{\bf E}|<E_c$ regime. In the case of M87, with a mass of a few $10^9~M_{\odot}$, the repetition timescale is $0.68$ d in the polar direction, with a quanta of $\mathcal{E}\sim 10^{45}$~erg \citep{2021A&A...649A..75M}}


\section{Conclusions}\label{sec:14}

GRB 190114C has offered already the possibility of testing different Episodes of the BdHN \textcolor{black}{I} sequence by a time-resolved spectral analysis \textcolor{black}{\citep{2021MNRAS.tmp..868R}}; \textcolor{black}{the $\nu$NS}-rise \textcolor{black}{[Becerra et al. in preparation],} the formation of the BH \textcolor{black}{triggering} the UPE phase \textcolor{black}{and} the associated GeV emission (see Fig.~\ref{fig:data}), the formation of the cavity \citep{2019ApJ...883..191R}, the long-lasting emission in the x-ray afterglow \textcolor{black}{from the spinning $\nu$NS \citep{2020ApJ...893..148R}, } and in the GeV \textcolor{black}{emission from the newly-formed BH \citep{2021A&A...649A..75M}}. 

\textcolor{black}{The long lasting GeV radiation, with a luminosity following a power-law of $ L_{\rm GeV} = \left(7.75\pm 0.44\right) \times 10^{52}~t^{-(1.2\pm 0.04)}$~erg~s$^{-1}$, has been shown to originate from the extraction of the rotational energy of a Kerr BH in a sequence of discrete ``blackholic quanta'' emission \citep{2020EPJC...80..300R}. This process occurs in an \textit{inner engine}, which is composed of an uniform magnetic field aligned with the rotation axis of the Kerr BH described by the Papapetrou-Wald solution \citep{1966AIHPA...4...83P,1974PhRvD..10.1680W} and immersed in a very low density fully ionized plasma with density as low as $10^{-14}$ g cm$^{-3}$ \citep{2019ApJ...883..191R,2020EPJC...80..300R,2021A&A...649A..75M,2021MNRAS.tmp..868R}.}

\textcolor{black}{One of the main results has been the concept of effective charge $Q_{\rm eff}$, given by Eq.~\ref{eq:EFCH} driving the acceleration process in the \emph{inner engine}. }

The most unexpected result has been the discovery of hierarchical structure in the time-resolved spectral analysis \textcolor{black}{on ever-decreasing time scales} of \textcolor{black}{the UPE phase of} GRB 190114C \textcolor{black}{by \citet[][]{2019arXiv190404162R} and here updated \textcolor{black}{in section \ref{sec:heir}}. There, we have determined} the spectral properties and luminosities during and \textcolor{black}{after} the UPE phase, of the MeV emission observed by \textit{Fermi}-GBM, and of the GeV emission observed by \textit{Fermi}-LAT.

A new arena \textcolor{black}{is open} in this article linking the macroscopic hierarchical structure \textcolor{black}{of the UPE phase} to a microphysical sequence of discrete elementary events \textcolor{black}{in a QED regime}. 

For the first time, we have \textcolor{black}{here} approached the energy extraction process from a Kerr BH by the general relativistic QED process occurring in the \emph{inner engine}.

 
\textcolor{black}{We have assumed} that the electric field of the \emph{inner engine}  \textcolor{black}{operates in an} overcritical $|{\bf E}|>E_c$ \textcolor{black}{during} the UPE phase, and \textcolor{black}{in an} undercritical $|{\bf E}| < E_c$ just after the end of the UPE \textcolor{black}{phase}. A sharp separatix both in the theoretical treatment and in the observational properties of these two domain\textcolor{black}{s} are evidenced.  



The main result of this article is to have \textcolor{black}{compared and contrasted the} two different processes \textcolor{black}{for} explaining the MeV and GeV emissions of GRBs. 

The first process, \textcolor{black}{originating the MeV radiation,} is dominated by the vacuum polarization originating from the overcritical field in the UPE phase. \textcolor{black}{The overcritical field generates an initially optically thick $e^+~e^-~\gamma$-baryon plasma, which self-accelerates until  reaching  the  point of transparency, a PEMB pulse. Typical values of $\Gamma\sim 100$ guarantee the avoidance of the compactness problem \citep[][]{1975NYASA.262..164R,2004RvMP...76.1143P} in the UPE phase.} \textcolor{black}{We have shown that t}he magnetic field $B_0$ \textcolor{black}{ keeps a constant value during the UPE phase} of order of $\sim10^{14}$~G and reduces to $3.9 \times 10^{10}$~G after the UPE phase. 


The second process, \textcolor{black}{originating} the GeV emission, is based on the classical ultrarelativistic electrodynamics generated from the electrons injected in the magnetic field emitting synchrotron radiation \textcolor{black}{close to the BH horizon}, in selected energies with specific pitch angle dependence, see Fig.~9 in \citet{2019ApJ...886...82R}.

Both these processes originate from the rotational energy of the \textcolor{black}{Kerr} BH acting on a uniform magnetic field, aligned with the BH rotation axis, \textcolor{black}{within Papapetrou-Wald solution}. 

The result\textcolor{black}{s} presented here \textcolor{black}{were} expected since fifty years when the Christodoulou-Hawking-Ruffini mass-energy formula of the BH  \citealp{1970PhRvL..25.1596C,1971PhRvD...4.3552C,1971PhRvL..26.1344H}, as well as some of the pioneering works, using the vacuum polarization process of a BH, were established \citep{1976PhRvD..14..332D,1998astro.ph.11232R}. They were followed by fundamental contributions on the self-acceleration process of the $e^+~e^-~\gamma$~optically thick plasma, PEMB pulses \citep{1998astro.ph.11232R,1999A&A...350..334R, 2000A&A...359..855R}, and by the concepts of \textit{dyadosphere} and \textit{dyadotorus} \citep[see][and references therein]{2010PhR...487....1R}, \textcolor{black}{which are the fundamental conceptual framework of this article. The revival of these concepts, as we explained in this article, has been made possible by the fundamental introduction of ``\emph{the effective charge}'' overcoming the concept of a net charged BH and fulfilling, nevertheless, all the necessary electrodynamical process of an electrically charged BH.}



The fact that all the properties of GRB 190114C have been confirmed to occur in GRB 130427A, GRB 160509A and GRB 160626B; see \citep{2019arXiv191012615L}, allow us to extend and apply the analysis here performed for the \emph{inner engine} , generally to all BdHNe I.

This has introduced a radical change by modifying the traditional energetic arguments based on the gravitational binding energy of \textcolor{black}{massive particles} geodesics, \textcolor{black}{following a classical electrodynamics process} in the Kerr metric occurring at very high density. Indeed, in \cite{2019ApJ...883..191R}, it has been shown how this \emph{inner engine}  operates \textcolor{black}{most efficiently} in a cavity in presence a very tenuous ionized plasma with density of $10^{-14}$~g~cm$^{-3}$ \textcolor{black}{ following a classical electrodynamics process. The \emph{inner engine} equally works at high densities of the PEMB pulses in the quantum electrodynamics process. In both processes, t}he fundamental energetic role is being played by the rotational energy of the Kerr BH, which is converted by associated classical and quantum ultrarelativistic acceleration processes into the observed multi-wavelength energy emissions and UHECRs.  The application of \textcolor{black}{the} classical work of the innermost stable circular  orbit (ISCO) \textcolor{black}{of massive particles around Kerr BH}, introduced in \citet{Ruffini:1971bza}, has been superseded in this new approach. \textcolor{black}{The concepts of the dyadosphere and \citep{1998astro.ph.11232R,1999A&A...350..334R, 2000A&A...359..855R} and dyadotorous \citep{2009PhRvD..79l4002C}s are the fundamental ones in this new electrodynamical scenario.}

The most important result \textcolor{black}{in this paper has been the understanding the role of } hierarchical structure discovered in the time-resolved spectral analysis of the UPE phase, \textcolor{black}{finally explained by their} underlying quantum nature.

\textcolor{black}{This long march was started by the intuitions announced in  \citet{2019arXiv190404162R}. They have been here expanded and approached in their theoretical implication in this article. Although the motivations were clear, their detailed comprehension has needed further work which is here presented. We are ready to look at the implications of these results.}

%
%
%
%
%



Thanks to the observation of GRB 190114C, which is by far the most complex fundamental physical system ever approached in Science, a new scenario is now open. The most unique complexity of BdHNe\textcolor{black}{,} their enormous energy emitted in an observer homogeneous Universe, see e.g., \citet{1988A&A...190....1R}, \textcolor{black}{ and the special quantum ans classical electrodynamics nature of their radiation} make us wonder about the role GRB may play in the appearance of life in the Universe \textcolor{black}{\citep{2015ARep...59..469C}}. This new overarching conceptual description appears to be in sight thanks to the observation of GRB 190114C.

\acknowledgements
\textcolor{black}{We thank the referee for important remarks which have improved the presentation of our results.} We are grateful to Prof. Roy Kerr for discussion on the energy extraction process from Kerr BH, to Prof. Sang Pyo Kim for fruitful discussion about overcritical field and Schwinger pair production in such a field \textcolor{black}{and to Prof. Narek Sahakyan for studying the extension of these results from GRBs to AGNs}. 



\providecommand{\noopsort}[1]{}\providecommand{\singleletter}[1]{#1}%

\end{document}